\documentclass[preprint, sort, compress]{elsarticle}
\usepackage[utf8]{inputenc}
\usepackage{graphicx}
\usepackage{bm}
\usepackage{amsmath}
\usepackage{amssymb}
\usepackage{dirtytalk}
\usepackage{soul}
\usepackage{array}
\usepackage[hidelinks]{hyperref}
\usepackage[indent]{parskip}

\hypersetup{ breaklinks=true}
\begin{document} 

\title{Network Risk Estimation: A Risk Estimation Paradigm for Cyber Networks}

\author[1]{Arda Bayer\corref{cor1}}
\ead{Arda.Bayer@rice.edu}

\author[2]{David Maluf}
\ead{dmaluf@cisco.com}

\author[1]{Behnaam Aazhang}
\ead{aaz@rice.edu}

\cortext[cor1]{Corresponding author}
\affiliation[1]{organization={Rice University},
addressline={6100 Main St},
postcode={77005 TX},
city={Houston},
country={U.S.}}

\affiliation[2]{organization={Cisco},
addressline={170 West Tasman Drive},
postcode={95134 CA},
city={San Jose},
country={U.S.}}

\date{January 2025}

\begin{abstract}

    Cyber networks are fundamental to many organization's infrastructure, and the size of cyber networks is increasing rapidly. Risk measurement of the entities/endpoints that make up the network via available knowledge about possible threats has been the primary tool in cyber network security. However, the dynamic behavior of the entities and the sparsity of risk-measurable points are limiting factors for risk measurement strategies, which results in poor network visibility considering the volatility of cyber networks. This work proposes a new probabilistic risk estimation approach to network security, NRE, which operates on top of existing risk measurements. The proposed method NRE extracts relationships among system components from the network connection data, models risk propagation based on the learned relationships and refines the estimates whenever risk measurements are provided. In this work, \textit{(i)} the risk estimation scheme is proposed, \textit{(ii)} an application of quantitative risk estimates is devised, \textit{(iii)} descriptiveness of the risk estimates are compared to a pure risk measurement alternative and \textit{(iv)} low computational complexity of the proposed method is illustrated capable of real-time deployment. The proposed method, NRE, is ultimately a quantitative data-driven risk assessment tool that can be used to add security aspects to existing network functions, such as routing, and it provides a robust description of the network state in the presence of threats, capable of running in real-time.        
\end{abstract}


\maketitle

\section{Introduction}
\label{sec:intro} 
\begin{figure}[ht]
    \centering
    \includegraphics[width= \linewidth]{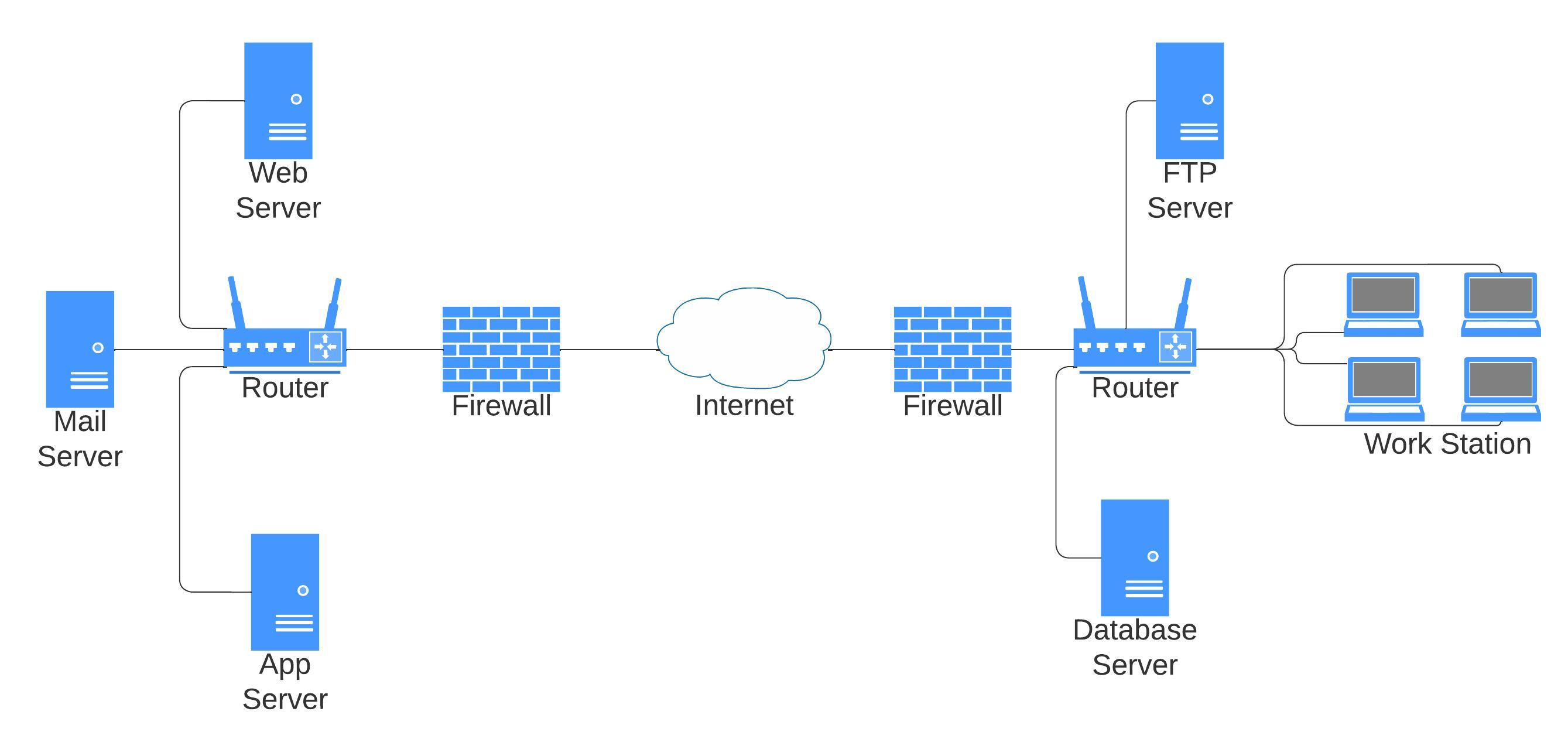}
    \caption{An enterprise network \cite{chakravarthiamonitoring} consisting of users, servers and routers. The interconnection of these entities makes up the network topology, which indicates existing direct communication channels. The network topology of a cyber network is the starting point of \textit{Risk Estimation}.}
    \label{fig:ent_net}
\end{figure}

Cyber networks play a crucial role in the functioning of many organizations. A cyber network, or network as short, is essentially any interconnection of individual devices referred to as \say{entities} or \say{endpoints} that communicate with each other or make their communication possible. Typical examples of networks are \textit{(i)} domestic networks (smartphones, TVs, lighting systems, etc.), \textit{(ii)} enterprise networks (consisting of users, servers, and routers, as depicted in Figure \ref{fig:ent_net}), \textit{(iii)} industrial networks (e.g., sensors that gauge machine performance and communicate it to a Command and Control Center), and also \textit{(iv)} networks of devices at hubs such as airports and malls. Recent advancements in technologies such as wireless, the Internet of Things (IoT), and increased adoption of cloud computing have rapidly increased the effective size of the ultimate cyber network, where previously mentioned networks are frequently brought together. Accordingly, studies indicate that the size of the ultimate cyber network, coupled with the cloud, is increasing rapidly and poses a challenge when assessing the network regarding security \cite{cisco_2023}.

Risk assessment is the primary tool of cyber network security against countless threats \cite{cloud_risk_assess, albakri2014security, ou2011quantitative, ralston2007cyber}. Cyber networks make up the infrastructure of many organizations and institutions, making them attractive targets for malicious attacks. A typical attack targeted at a cyber network works from the most vulnerable endpoint to its target. Attackers strive to exploit any weaknesses, and defenders try to patch the inherent vulnerabilities of their system components \cite{maluf2018trust}. Against a wide variety of vulnerabilities, the status quo of cyber network security has been \say{risk measurements}, with a unifying notion of risk for a selection of vulnerabilities. 

The term \textit{risk}, sometimes called vulnerability score \cite{ou2011quantitative, ramos2017model}, is used to quantify the expected loss if a security breach were to occur while considering the likelihood of that breach \cite{maluf2018trust, ralston2007cyber, quirc}. The current defenses for cyber networks measure the risks in a principled way by prioritizing vulnerabilities based on their associated costs and leveraging existing knowledge about the threat to estimate the likelihood of that threat \cite{ou2011quantitative}. Accordingly, a cyber security system that gathers information about entities and quantifies network vulnerability is a risk measurement system that gauges the cyber network's risk through available entities. Some examples of risk measurement tools are antivirus software, intrusion detection systems (IDSs), and network endpoint analytics tools \cite{cisco_2023}.

Since entities are ultimately responsible for the risk at the network level, the risk measurement practice can be viewed as a way to probe the vulnerability of the cyber network through available entities. However, even with the ideal risk measurements, in practice, the risk-measurable entities are generally a small portion of the network, and the growing size of the network, along with changing entity behavior, are currently limiting factors of risk measurement practice in cyber network security \cite{cloud_risk_assess, ou2011quantitative}. 

In this work, a scalable and data-driven \textit{risk estimation} paradigm, referred to as \say{Network Risk Estimation (\textit{NRE})}, is designed to address these challenges. The proposed system works within a probabilistic framework where a risk distribution on entities is estimated using (1) the connection data and (2) the \textit{risk measurements} whenever and wherever provided. To our knowledge, this is the first work to model risk propagation based on entity relationships. The premise of the \textit{NRE} is that the threat will propagate from the entity that is the threat's source to the entities associated with it. This premise lays out the spatial -across entities- and temporal -across time- dynamics of the risk propagation model of NRE. The result is an estimated probability distribution of risks among entities over time.

\begin{figure}

  \centering
  \includegraphics[width=0.8\linewidth]{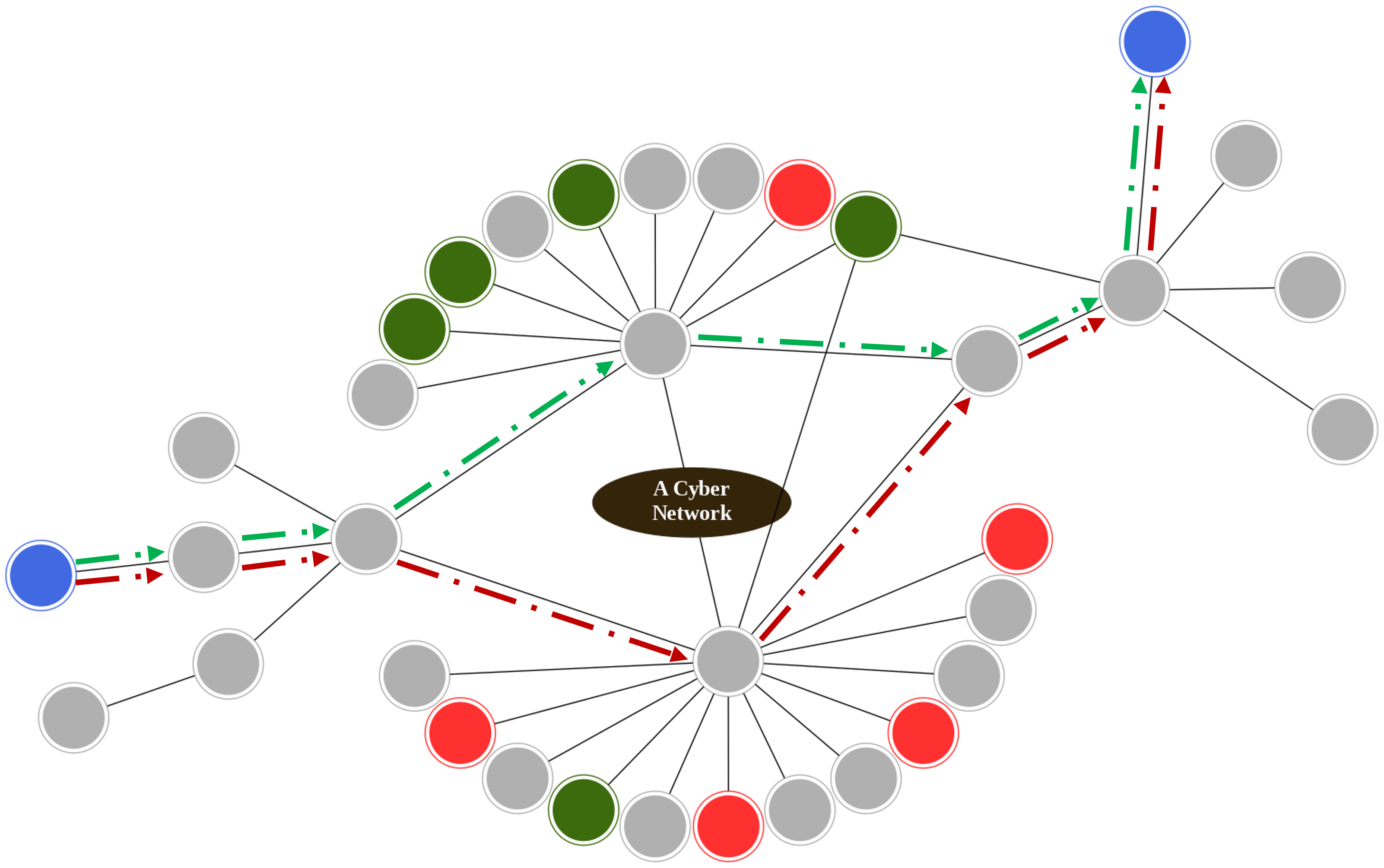}  

\caption{ An application of risk estimation on a cyber network. The graph gives the network topology where each node is an entity, and an edge indicates an allowed communication channel. Red entities are measured to have relatively higher risks, green entities have low measured risks, and blue entities desire to communicate. Two possible routes are depicted via green and red dashed arrows. For safe routing, the green path is chosen over the red one since measured entities in the vicinity have less risk.}
\label{fig:net}
\end{figure}

The probabilistic risk estimates given by the proposed system can be used to \textit{(a)} assess the network security quantitatively and \textit{(b)} add security aspect to network-related functions. For instance, entities with high risks can be identified as weak points in the network. Alternatively, as a more elaborate network management action, the safety aspect can be added to packet routing as illustrated in Figure \ref{fig:net}, which is briefly discussed in Section \ref{sec:application}. Our contributions are \textit{(i)} entity relationship modeling from available connection data, \textit{(ii)} partitioning of the network via discovered entity relationships, \textit{(iii)} explicit probabilistic estimation of entity risks, \textit{(iv)} and an application of these estimates in network management. 

The rest of the paper is organized as follows. Section \ref{sec:related_work} briefly summarizes the related work in risk analysis, risk measurement approach to vulnerability assessment, and network modeling for intrusion detection. Section \ref{sec:problem_statement} formally states the problem of interest. In Section \ref{sec:method}, the details of the proposed method are provided. Finally, Section \ref{sec:results} presents (1) the overall risk estimation process and the results on a public dataset, (2) a direct application of the produced risk estimates, (3) a particular proof of concept experiment designed for comparison with the alternative method, and (4) the running time complexity analysis.

In addition, the implementation of the work that is described in this paper is open source and can be accessed at \href{https://github.com/ab126/NRE}{https://github.com/ab126/NRE}.

\section{Related Work}
\label{sec:related_work}
The subject of risk assessment is well-established, and it is common to many engineering designs \cite{risk_analysis}. There exist standards for risk assessment regarding network security that have been set from an organizational standpoint. ISO 31000 \cite{iso31000:2018} lays out general guidelines for organizations of any kind for the purpose of assessing the risks that may hinder the organization's normal operation. These guidelines set the standard for risk management and give general instructions to deal with organizational risks where risk is essentially defined the same as in this work; the likelihood of a threat times the impact. However, it lacks the specificity to be directly applied to a cyber network. For example, cloud computing is a substantial cyber network instance that is extremely dynamic and volatile. For the cloud, these standards were criticized for being unable to cover the cloud environment's dynamic nature, and a proactive method was suggested \cite{cloud_risk_assess}. In accordance with this shortcoming, some qualitative and quantitative risk assessment work has been dedicated to cloud computing risk assessment \cite{cloud_risk_assess, albakri2014security, quirc, sendi_cloud}. Among these studies, \say{Quantitative Impact and Risk Assessment Framework } \cite{quirc} presents a risk assessment framework that relies on the consensus of an expert team on the impacts of various risks and security reports in estimating the likelihood which is used to get a measurement of risks. Its main drawback is that it fails to be adaptive due to the reliance on expert discussion and lack of use of real-time network data. Another work that was aimed to quantify risks in a cloud environment is \say{Cyber Supply Chain Cloud Risk Assessment} \cite{cloud_risk_assess}, which similarly follows the ISO 31000 guidelines and outlines a quantitative risk assessment framework with an emphasis on the supply chain side of the cloud. This work puts forth the idea of making use of the topological information in supply-side components of the cloud service and uses probability distributions to model the uncertainty inherent in calculating the risks. However, this risk measurement approach is also not fully data-driven in the sense that it still requires cloud service provider stakeholders' knowledge and intervention to produce quantitative risks for the whole network. The authors also point out the need for a \say{dynamic \& adaptive} risk assessment technology for the cloud environment as future needs \cite{cloud_risk_assess}.

The studies mentioned so far can be treated as direct risk measurement practices on cyber networks where the outcome is the risk values. However, there are also works done on indirect risk measurements that deduce the likelihood part of the risk for a particular threat. The most common studies of this category are Intrusion Detection Systems (IDSs) that are developed for various cyber-physical networks against certain threats. For the intrusion detection problem, there exist labeled network data, called herein as \textit{connection data}, which consists of the collection of communication sessions between pairs of entities called \textit{flows}. A collection of descriptors called \textit{flow features} can be extracted from raw bits making up the flows \cite{cicids2017}, after which the intrusion detection problem becomes a classification problem on the flows. An example of such an IDS is given by \textit{Wu et. al.} where the flows are converted into doc-matrix format based on the presence of particular intervals of flow features, which are then used in the classification of flows \cite{wu2022intrusion}.

Another approach for modeling intrusion detection has been the incorporation of network topology into the detection task via graph structure. Entity relationships gathered from connection data have previously been summarized in a \say{Communication Graph} \cite{nagaraja2014botyacc} where nodes are the entities and edges are present if two respective nodes have participated in a flow. In their work, after obtaining the communication graph, the intrusion detection problem boils down to an edge partitioning problem on the communication graph, which has been solved by transforming it to a canonical node partitioning problem on the dual graph \cite{nagaraja2014botyacc}. Variants of the communication graph have been studied in intrusion detection with the same idea of explicitly using entity relationships (see \cite{dias2020big} for a survey). For instance, different graph neural network architectures were designed by \textit{Chang et. al.} to make the best use of both structural information in the graph, the entity relationships, and the information in the flow descriptors \cite{chang2021graph}.

Despite the vast modeling perspectives, these studies from the intrusion detection literature are essentially \textit{risk measurements} for particular threats, such as \say{Distributed Denial of Service} or \say{Brute Force} attacks \cite{cicids2017}, from risk assessment \cite{risk_analysis} point of view. Consequently, as mentioned in Section \ref{sec:intro}, this approach to cyber network security suffers from limited risk-measurable entities, the dynamic nature of the network in terms of changing entity behavior, and the addition of new entities.

\section{Problem Statement}
\label{sec:problem_statement}
In the security setting, network assessment is done in terms of risk measures  \cite{risk_assess1, maluf2018trust, quirc}. Once the risks of individual threats are identified, defensive measures can be taken \cite{quirc, sendi_cloud}. As an example, if the greatest risk against the network is an exploit due to an older version of the software, a software update can be administered to the respective entities in the network. Following the same idea about the network-level risks, the risk of an entity for a particular threat is proposed in this work as the \textit{likelihood of the entity being compromised} times \textit{the impact of the same entity being compromised}. Here, the impact is the quantification of the cost if the unwanted scenario were to happen. An example the cost in dollar amounts if the respective entity was compromised to a virus. Ultimately, this definition of entity risk is along the same line as the conventional definition of risk used for the whole network \cite{iso31000:2018, maluf2018trust, ralston2007cyber, quirc} but at the entity level. 

Accordingly, the goal of this work is to estimate the risks of each entity in the network, given (1) \ul{some risk measurements on a set of entities} and (2) \ul{the connection data, the flows, that capture the activity in the network}. Phrased in the framework of probability theory, the problem tackled here is to find the probability distribution of entity risks given sparse risk measurements on some entities, red and green entities in Figure \ref{fig:net}, and connection data. It is implied that the risk measurements, whatever the source be, are mapped to the same \say{risk units} or scale according to the same standard. 

When risks are forged for every entity, network management becomes clear. For instance, given two entities that want to communicate, e.g., blue entities in Figure \ref{fig:net}, one can find the \say{most trusted path} to route the packets in between, which is referred to as safe routing (Section \ref{sec:application}). One way to do this is to minimize the expected maximum risk along the path over all possible paths between two entities, intuitively picking the green path instead of the red one in Figure \ref{fig:net}. This application of the proposed risk estimation method is illustrated in Section \ref{sec:application}. Another use case is that the \say{high risk} entities in the network can be immediately identified and can be used in devising new measurement strategies or in implementing safety rules such as quarantining.

\section{Method} \label{sec:method}
This section outlines the details of the proposed risk estimation method. The underlying premise assumed in this work is that the risk of entities will propagate through the network by the way entities influence each other. This is based on the fact that organized cyber-attacks target system vulnerabilities to gain privilege over entities \cite{attackgraphreview, ou2011quantitative}. It is assumed herein that during this process, the exploiter leaves a footprint in the connection data/flows, referred to as the influence relationship between involved entities. With this reasoning, the first part of the proposed work is dedicated to uncovering the way entities influence each other in the network, which is the basis for the \textit{functional connectivity graph} (Section \ref{sec:relationship_inference}). This step summarizes the discovered entity behavioral relationships as a weighted directed graph $\mathbf{F}$. In the second part, the calculated graph is used as a linear model for predicting risks, in which the risk estimates are refined with incoming measurements using a probabilistic framework (Section \ref{sec:risk_prop}). Finally, in Section \ref{sec:real_time_op}, real-time deployment is discussed. Although the proposed method is used in the context of cyber networks, it is generalizable to other dynamic networks for obtaining low dimensional latent variables regarding system state.

\subsection{Inference of Relationships}
\label{sec:relationship_inference}

For the cyber networks considered in this work, it is assumed that the connection data comprises distinct flows just like NetFlow, sFlow formats \cite{hofstede2014flow} where a \textit{flow} is defined as any communication between two endpoints in one session \cite{hofstede2014flow, sperotto2010overview}. These flows have multiple \textit{flow attributes} that give additional information about flows such as timestamp, flow duration, and number of packets in each direction and more \cite{cicflowmeter}.

Previously in the field of cyber-security, similar entity relationship graphs have been constructed as \say{\textit{communication graphs}} \cite{dias2020big, nagaraja2014botyacc} where two entities have an edge if there has been at least one flow observed between them. Although the communication graph is a valid way to model the entity relationships, we here present a more general way to devise the relationship graph, referred as the \textit{functional connectivity graph}, encompassing the communication graph but also capable of capturing more abstract dependencies among entities.

Given the collection of flows referred as the \textit{connection data}, the relationship among entities is analyzed in this work with respect to a single \textit{connection parameter} derived from flow attributes. Each connection parameter captures a distinct aspect of the flows. For instance, \say{Number of Packets Received} is a connection parameter derived from \say{Number of Packets in the Forward Direction} and \say{Number of Packets in the Backward Direction} flow attributes. It represents the inward volume of flows for an entity. Ultimately, the connection parameter models the aspect of entity behavior for the risk prediction process. In this work, a few straightforward choices of connection parameters have been put forth. A summary of relevant connection parameters is given in Table \ref{table:conn_param}.

\begin{table}[!ht]
\caption{A Summary of Connection Parameters that are used. Each connection parameter captures an aspect of flows involving an entity, and it is derived from flow attributes in connection data. \textit{Aggregation Method} refers to the way flows are aggregated in the synchronization step (Section \ref{sec:synch}). The flow attributes are based on \textit{Lashkari et al.} cite{cicflowmeter}.}
\begin{tabular}{||>{\centering\arraybackslash} m{3.5cm} | >{\centering\arraybackslash} m{4.5cm} | >{\centering\arraybackslash} m{2.5cm}||} 
 \hline
\textbf{ Connection Parameter} & \textbf{Description} & \textbf{Aggregation Method} \\ 
 \hline

 Activation & A binary indicator denoting if an entity is sending or receiving any packets & N/A\\ 
 \hline
 Active Time & Mean time an entity was active before becoming idle & Average \\ 
 \hline
 Flow Duration & Total duration of the flows & Total\\ 
 \hline
 Flow Speed &  Average bytes per second & Average\\ 
 \hline
 Header Length &  Average bytes used in packet header in forward direction & Average\\ 
 \hline
 Idle Time &  Mean time the flows were idle before becoming active & Average\\ 
 \hline
 Number of Active Packets &  Total number of packets with at least 1 byte of packet payload in forward direction & Total\\ 
 \hline
 Number of Packets Received &  Total number of packets in the backward direction & Total\\ 
 \hline
 Number of Packets Sent & Total number of packets in the forward direction & Total\\ 
 \hline
 Packet Delay &  Average of packet Inter Arrival Time (IAT) in forward directions & Average\\ 
 \hline
 Packet Length & Average size of packets in forward direction & Average \\ 
  \hline
 Port Number & The port number used by the entity for the flow & Last \\ 
    \hline
 Protocol & Flow's Internet Protocol Number (e.g. TCP:6, UDP:17, etc.) & Last  \\ 
 \hline
 Response Time & Average of packet Inter Arrival Time (IAT) in backward direction & Average \\ 
 \hline
\end{tabular}

\label{table:conn_param}
\end{table}

Once a connection parameter is chosen, discrete-time signals are formed for each entity. This process requires a synchronization step described in Section \ref{sec:synch}. Then, in Section \ref{sec:graph}, the method used for obtaining the functional connectivity graph is presented. Section \ref{sec:forget} and Section \ref{sec:partition} discuss some of the optional but recommended steps taken to optimize the graph, which will be used as the essential model for risk prediction.

\subsubsection{Synchronization} \label{sec:synch}
A flow can start and end anytime in the network. Consequently, the observations about entities in connection data are asynchronous naturally. For modeling entity behaviors, working with synchronous discrete-time signals is easier. Accordingly, all the flows in a given time interval are aggregated with a custom choice of window size, yielding synchronous signals for entities.

\begin{figure}[ht]
    \centering
    \includegraphics[width= 0.85\linewidth]{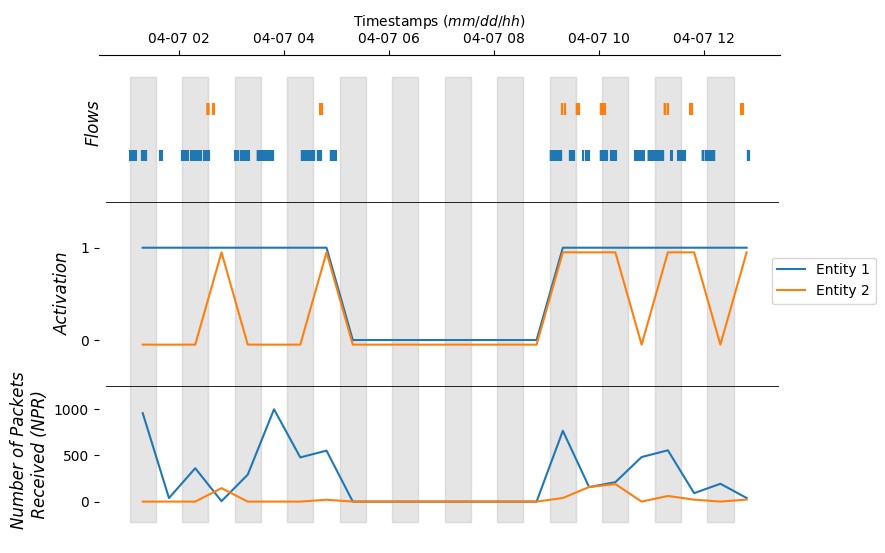}
    \caption{An illustration of the synchronization process involving two entities. Shaded and unshaded regions denote consecutive synchronization time windows. \protect\textit{Flows}: Flow history of two arbitrary entities. Each tick corresponds to a flow in connection data. \protect\textit{Activation}: The connection parameters that lead to a signal that depicts the activities of respective entities. \protect\textit{Number of Packets Received (NPR)}: Another connection parameter that grants a signal indicating the number of packets an entity receives. Ultimately, the obtained signals are defined per entity and are synchronous.}
    \label{fig:prewindow}
\end{figure}

For a chosen \textit{synchronization window size} denoted by $\delta$ (e.g., \textit{2 s}), flow attributes for each flow in a time window are aggregated by either summing, averaging, or taking the last value depending on the connection parameter that is being used (see Table \ref{table:conn_param}). As an example, if it is desired to build the entity-relationship model by the number of packets an entity receives, the \textit{\say{Number of Packets Received}} \textit{(NPR)} signal for each entity can be formed by counting the total number of packets each entity received per time-window. Similarly, if the objective is to model the flow speed, the \textit{\say{Flow Speed}} signal can be computed as the mean flow speed of the flows created by each entity. This process is summarized in Figure \ref{fig:prewindow}, where flows of two entities are depicted on the top row, and the constructed synchronous signals for two distinct connection parameters are given below. Here, the \say{Activation} is another connection parameter that can be used to deduce if two entities have sent packets coincidentally or not. Synchronous samples of these signals are then used to quantify influences among entities.

\subsubsection{Functional Connectivity Graph from Connection Data} \label{sec:graph}
After constructing synchronous signals per entity for a given connection parameter, the next step is to infer the underlying relationships among pairs of entities. The extent of entity A influencing entity B is quantified solely from the constructed signals of entity A and B with a choice of an influence measure. The end goal of this step is \textit{the functional connectivity graph} which is a weighted directed graph $G=(V, E, W)$ where $V$ is the set of entities, $E$ is the set of edges where an edge $e=(v_i,v_j)$ represent the influence of entity $v_i$ on $v_j$ and the respective non-negative weight $w_{ij} \geq 0$  quantify how strong the influence is for respective edge. 

\begin{figure}
    \centering
    \begin{minipage}{.45\linewidth}
    \centering
    \includegraphics[width= \linewidth]{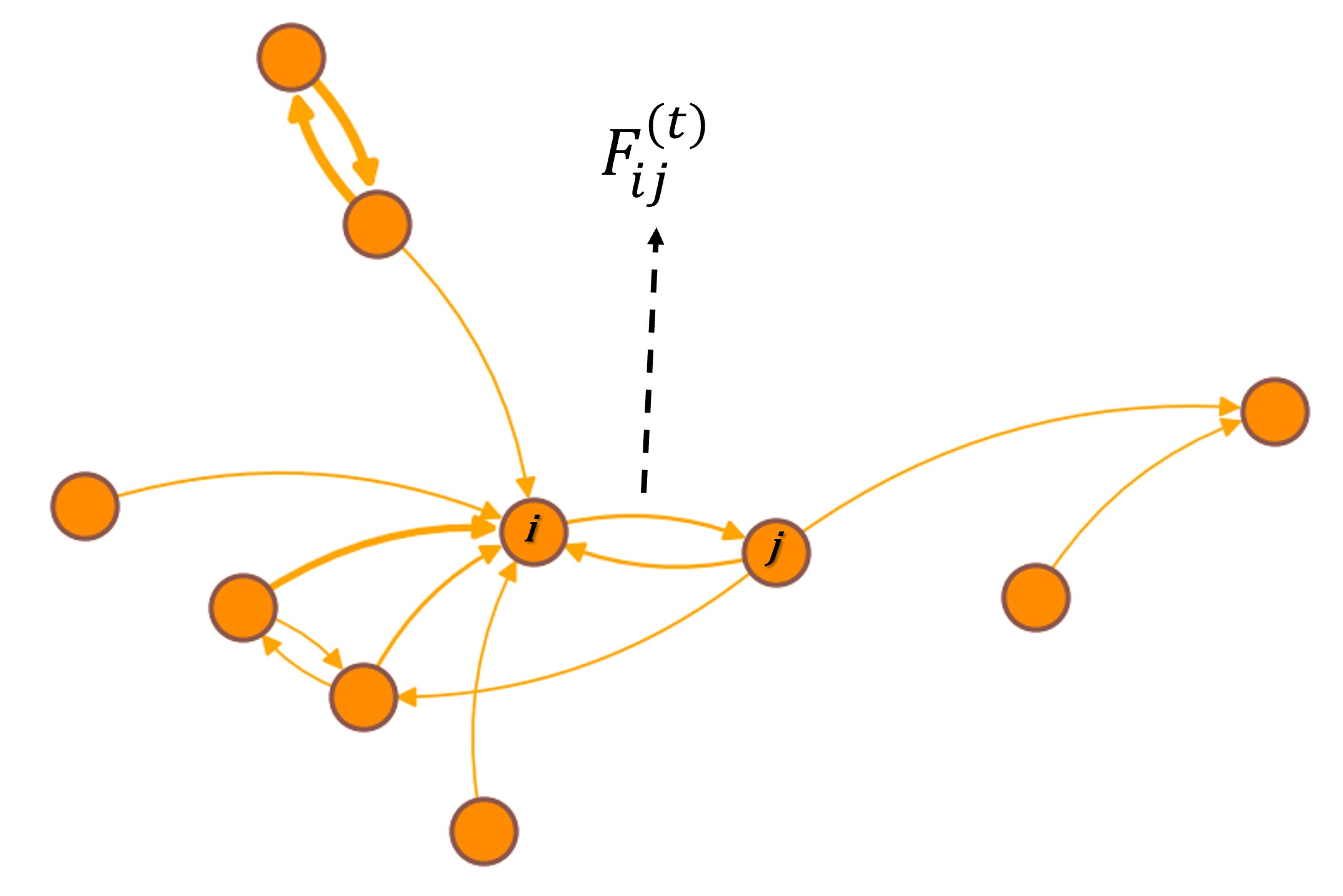}
    \textit{(a)} $\mathbf{F}^{(t)}$
    \end{minipage}
    \begin{minipage}{.45\linewidth}
    \centering
    \includegraphics[width=\linewidth]{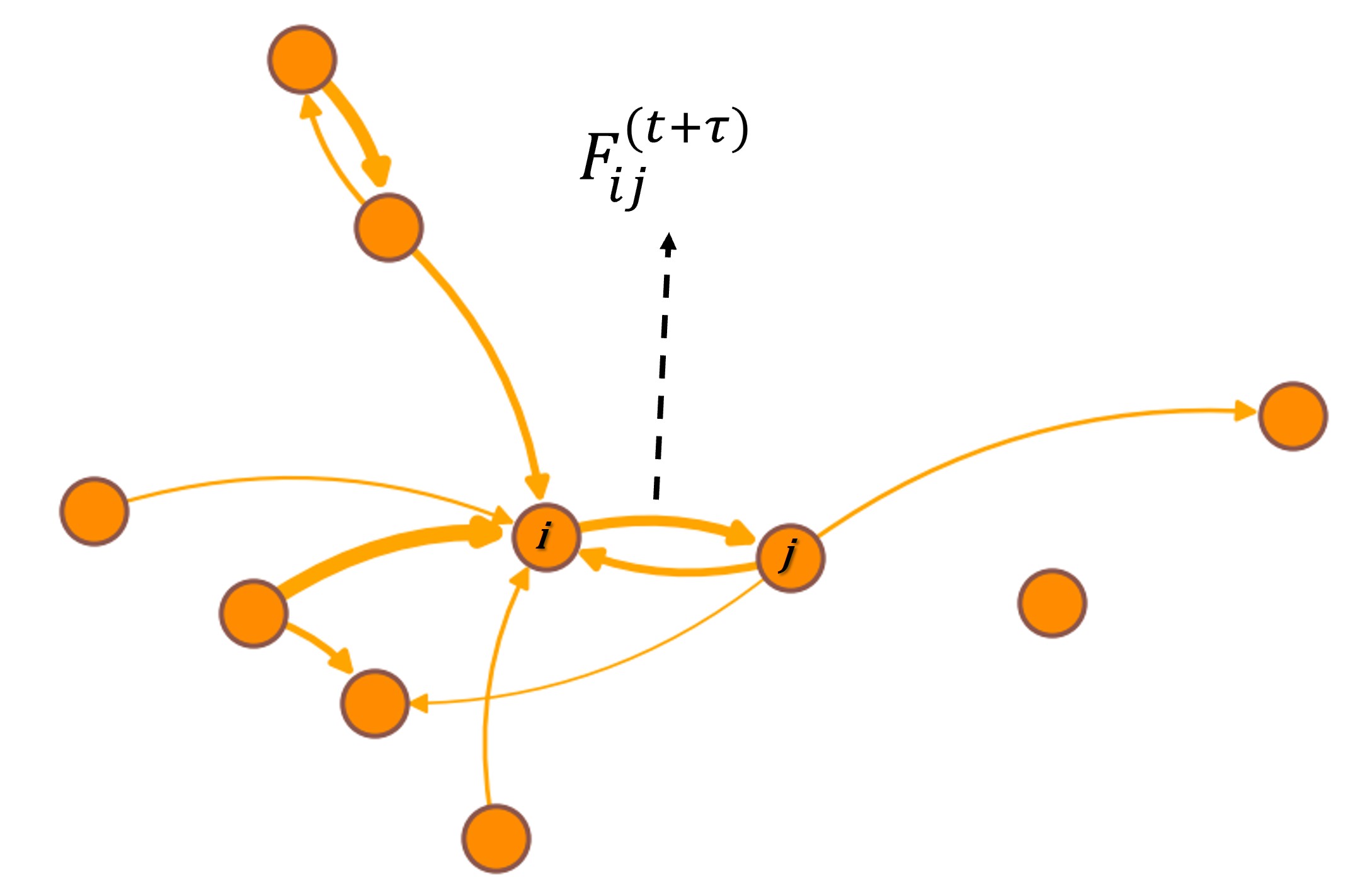}
    \textit{(b)} $\mathbf{F}^{(t+\tau)}$
    \end{minipage}
    \caption{A Functional Connectivity Graph $\mathbf{F}$ that manifests the relationships among entities at two different timestamps. \textit{(a)} $\mathbf{F}^{(t)}$ and \textit{(b)} $\mathbf{F}^{(t+\tau)}$. For each pair of entities $(i,j)$, the strength of influence $F^{(.)}_{ij}$ is calculated, which results in a weighted directed graph in general. The graph evolves with time as the connections happen.}
    \label{fig:rel_graph} 
\end{figure}

Once the entities are indexed, the graph $G=(V, E, W)$ is completely described by the respective \textit{weight matrix} denoted by $\mathbf{F}$ where the entries are the weights when the edge is present, and $0$ otherwise; $F_{ij} = \begin{cases}
  w_{ij}  &, (v_i, v_j) \in E \\
  0 &, (v_i, v_j) \notin E
\end{cases}$, as depicted in Figure \ref{fig:rel_graph}. The superscript $(t)$ is used to denote that the calculation of the graph was done for the time window $[t, t+\tau)$ where $\tau$ is the fixed \textit{graph window size}. Throughout this work, the weight matrix notation, $\mathbf{F}$, will be used to refer to the functional connectivity graph instead of the conventional set notation $G=(V,E,W)$.

Obtaining the underlying \textit{functional connectivity graph} out of observational data falls in the field of \say{Causal Discovery}, which comes with a few caveats for cyber networks. First, even if there was a unique underlying causal structure among entities in generating the connection data, the precise recovery of this structure is only possible up to a set of functional connectivity graphs that are indistinguishable without additional information \cite{glymour_causal}. Secondly and more critically, conventional causal discovery algorithms such as the Peter-Clark algorithm \cite{glymour_causal} and their variants scale terribly with the number of nodes/entities. The \say{Causal Discovery} is typically employed for problems having no more than 15 nodes, whereas in this case, the number of devices/entities can easily be thousands. For these reasons, inferring the functional connectivity graph from signals obtained according to Section \ref{sec:synch} is simplified greatly for the cyber network to what is called simple \textit{association network} inference \cite{horvath2011weighted}, which results in an undirected graph $\mathbf{F}$. In doing so, each pair of nodes is studied independently, and the objective weights of this graph, $w_{ij}$, are computed using a simple similarity metric for signals on the entities. However, the risk estimation approach proposed here is suited for any functional connectivity graph inference method that might be more appealing for different scenarios.

The similarity metric that has been used for weight calculation is the \say{Magnitude of the Pearson Correlation Coefficient}. In quantifying the relationship, each sample of the preprocessed signal is assumed to be independent with an appropriate choice of window size in Section \ref{sec:synch}, and thus, sample-based estimation of the correlation coefficient is justified as a metric for the association. The resulting graph is undirected or equivalently graph weight matrix $\mathbf{F}$ is symmetric.

Let $ Y^{(t)}_i = \{y_{i,1}^{(t)}, y_{i,2}^{(t)}, ..., y_{i,N}^{(t)}\} $ denote the $N$ samples of the synchronous signal obtained as in Section \ref{sec:synch} for the entity $i$ in the time window $[t, t+\tau)$. Then the functional connectivity graph for this time window obtained via the \say{Magnitude of the Pearson Correlation Coefficient} estimator is a symmetric weighted graph whose weight matrix's entries are given by (\ref{eq:F_cov}) where $Cov(Y^{(t)}_i, Y^{(t)}_j)$ is the sample covariance, $s (Y^{(t)}_i) $ is the sample standard deviation and $\overline{y_{i} }^{(t)} $ is the sample mean. The edge weights are normalized such that $F^{(t)}_{ij} \in [0, 1]$ and $ F^{(t)}_{ii} = 1 $, for the functional connectivity graphs in the proposed risk estimation scheme. The consequence of this normalization is that an entity will maintain its risk, and the entities that influence it will add to its previous risk (see Section \ref{sec:risk_prop}). The scaling of the absolute values of the risks are handled later in Section \ref{sec:relief}.

\begin{flalign}
    F^{(t)}_{ij} &= |r_{ij}| = \frac{|Cov(Y^{(t)}_i, Y^{(t)}_j)|}{s (Y^{(t)}_i) \, s (Y^{(t)}_j)} 
    \label{eq:F_cov}\\
    Cov(Y^{(t)}_i, Y^{(t)}_j) &= \frac{1}{N-1}\sum_{k=1}^N (y_{i,k}^{(t)} - \overline{y_{i} }^{(t)})(y_{j,k}^{(t)} - \overline{y_{j} }^{(t)}) \notag\\
    s (Y^{(t)}_i) &= \sqrt{ \frac{1}{N-1}\sum_{k=1}^N (y_{i,k}^{(t)} - \overline{y_{i} }^{(t)})^2}\notag\\
    \overline{y_{i} }^{(t)} &= \frac{1}{N} \sum_{k=1}^N y_{i,k}^{(t)}\notag
\end{flalign}

The calculated functional connectivity graph is per time-window that has been chosen, and the resulting graph $\mathbf{F}^{(t)}$ is dynamic, and it most likely will change over time as illustrated in Figure \ref{fig:rel_graph}. Although the inferred functional connectivity graph might resemble the network topology, i.e. the allowed communication channels between entities such as the one in Figure \ref{fig:net}, there is a crucial difference. Distant entities in the network topology might influence each other, resulting in a significant edge in the functional connectivity graph; however, there might be no direct edge in the network topology. For this reason, it is better to work with the functional connectivity graph from risk estimation point of view.

In forming the functional connectivity graph, different metrics such as Conditional Shannon Entropy, Mutual Information \cite{wang2015brain}, Directed Information \cite{quinn2011estimating}, or Graphical Directed Information \cite{young2021inferring} can be chosen. This preference would, in general, result in a weighted directed graph $\mathbf{F}^{(t)}$. However, the experiments presented in Section \ref{sec:results} are confined to the metric given by (\ref{eq:F_cov}), which leads to a symmetric graph weight matrix $\mathbf{F}^{(t)}$. One thing that must be considered as a result of using different similarity metrics is the scaling of risks. Having unbounded values in the entries will greatly affect the risk estimation step, Section \ref{sec:risk_prop}, thus we highly recommend normalizing the functional connectivity graph in one way or another. On top of the normalized functional connectivity graph $\mathbf{F}^{(t)}$, an additional mechanism to control the scaling of risks is designed in this work which is discussed in Section \ref{sec:relief}.

\subsubsection{Forget Factor}
\label{sec:forget}

In practice, it is desired for calculated functional connectivity graph $\mathbf{F}^{(t)}$ to have a \say{memory} such that the current connectivity graph carries over some of the relationships captured in previous time windows. This memory mechanism is modeled in the proposed work by (\ref{eq:forget_factor}) as:

\begin{equation}
    \mathbf{F}^{(t)}_{s} = (1-\rho_{f})  \mathbf{F}^{(t-\tau)}_{s} + \rho_{f} \mathbf{F}^{(t)}
    \label{eq:forget_factor}
\end{equation}

Here, $\rho_{f}$ is referred to as the \textit{forget factor}, and it controls the contribution of the past edge weight on the current functional connectivity graph. The value of $\rho_{f}$ is experimentally determined by validation on the training part of the dataset. The graph $\mathbf{F}^{(t)}_{s}$ is referred to as the \textit{smoothed graph} since it will change less dramatically with time and, it can be used in place of $\mathbf{F}^{(t)}$ from Section \ref{sec:partition} and onwards.

\subsubsection{Node Partitioning}
\label{sec:partition}
\begin{figure}
    \centering
    \includegraphics[width= 0.9\linewidth]{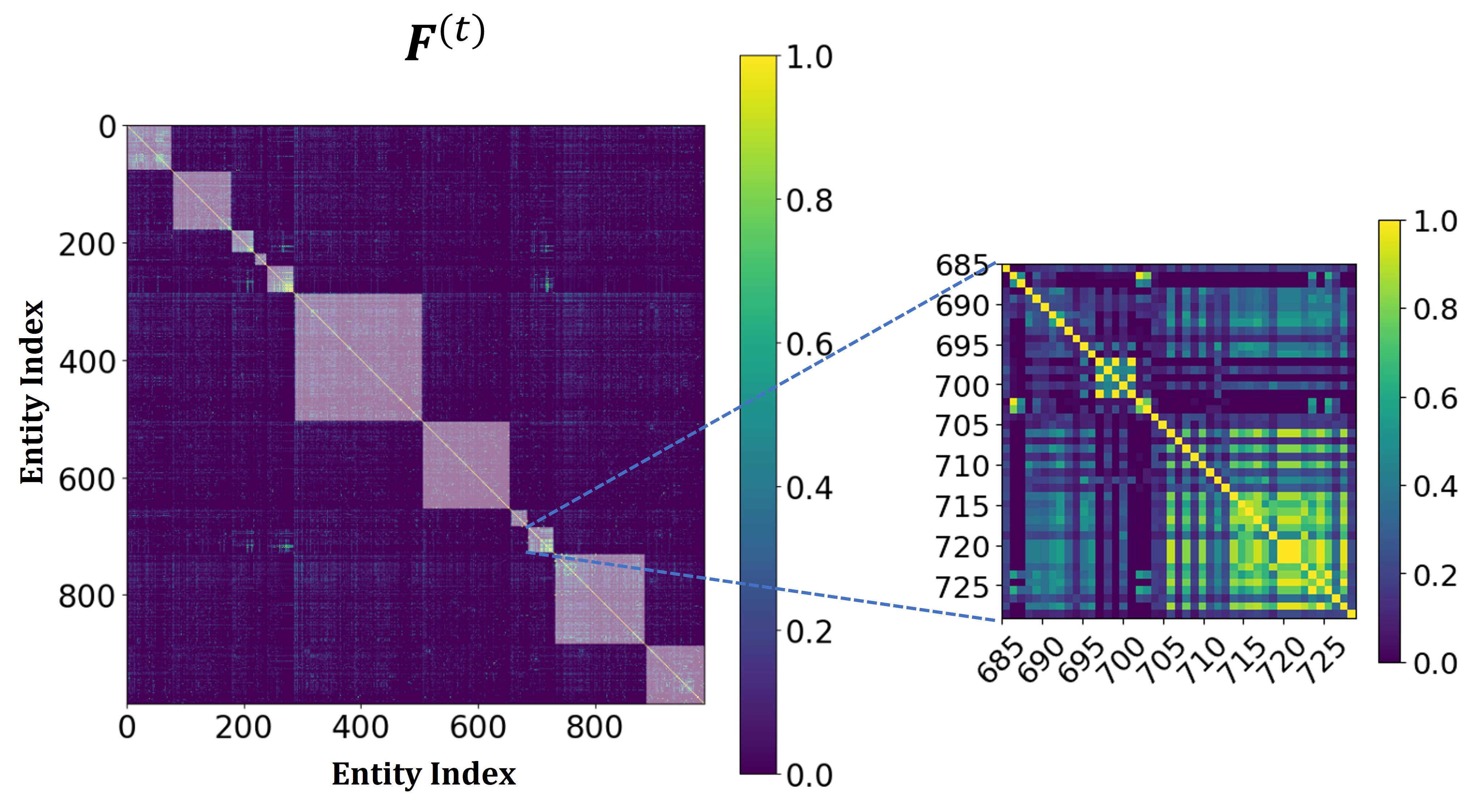}
    \caption{Entity groups detected via spectral partitioning algorithm. \textit{Left}: The matrix representation of the Functional connectivity graph $\mathbf{F}^{(t)}$ for the whole network where entities are enumerated on both axes. The $(i,j)$th entry, or $F^{(.)}_{ij}$, is the weight quantifying how much entity $i$ influences entity $j$. \textit{Right}: A sub-network of interest. Risk estimation is solved independently on the detected entity groups.}
    \label{fig:spec} 
\end{figure}

Occasionally, the calculated functional connectivity graph having thousands of entities will be sparse due to the local activity of entities. In those cases, it is much more desirable to break the problem into smaller subproblems by isolating functionally independent groups of entities via node partitioning and treating them independently for risk calculation (Figure \ref{fig:spec}). In practice, the separated entities will have small but non-zero edge weights between them. The node partitioning approximation essentially discards these inter-group edges, which are lost information. In order to reduce the approximation error, the node partitioning objective has been set to minimize the sum of the weights of the discarded edges, known as the \say{Ratio Cut} problem \cite{spectral}.

For the proposed method, the \textit{spectral partitioning} algorithm has been used as the preferred partitioning method, which solves the relaxed ratio cut problem where the sum of inter-group edges is minimized with a penalty against smaller groups \cite{spectral}. The partitioning can be applied until the entity groups are of the desired size, e.g., small enough to be processed in real-time. However, it must be noted that this is still an approximation, and each additional partitioning will increase the approximation error. Ultimately, spectral partitioning suggests a set of entities that can be studied independently with minimal error, which means that the risk prediction problem can be broken down into smaller, much more manageable problems in which the entities are well-connected.

\subsection{Risk Propagation Model} \label{sec:risk_prop}
The underlying assumption for the proposed risk propagation model is a linear model concerning how the risks propagate throughout the network. Suppose we have the risk estimates for each entity in the network at time $t$ represented by a column vector $\mathbf{x}_{t} \in \mathbb{R}^n$ where $n$ is the number of entities. The risk propagation model assumed here can be stated as \say{the increase in the risk of a given entity \textit{A}, is proportional to the risks of the entities that influence \textit{A} and to the extent they influence \textit{A}}. In other words, if \textit{A} is clearly associated with a high-risk entity, we expect \textit{A}'s risk to increase accordingly.

\begin{figure}
    \centering
    \includegraphics[width= 0.8\linewidth]{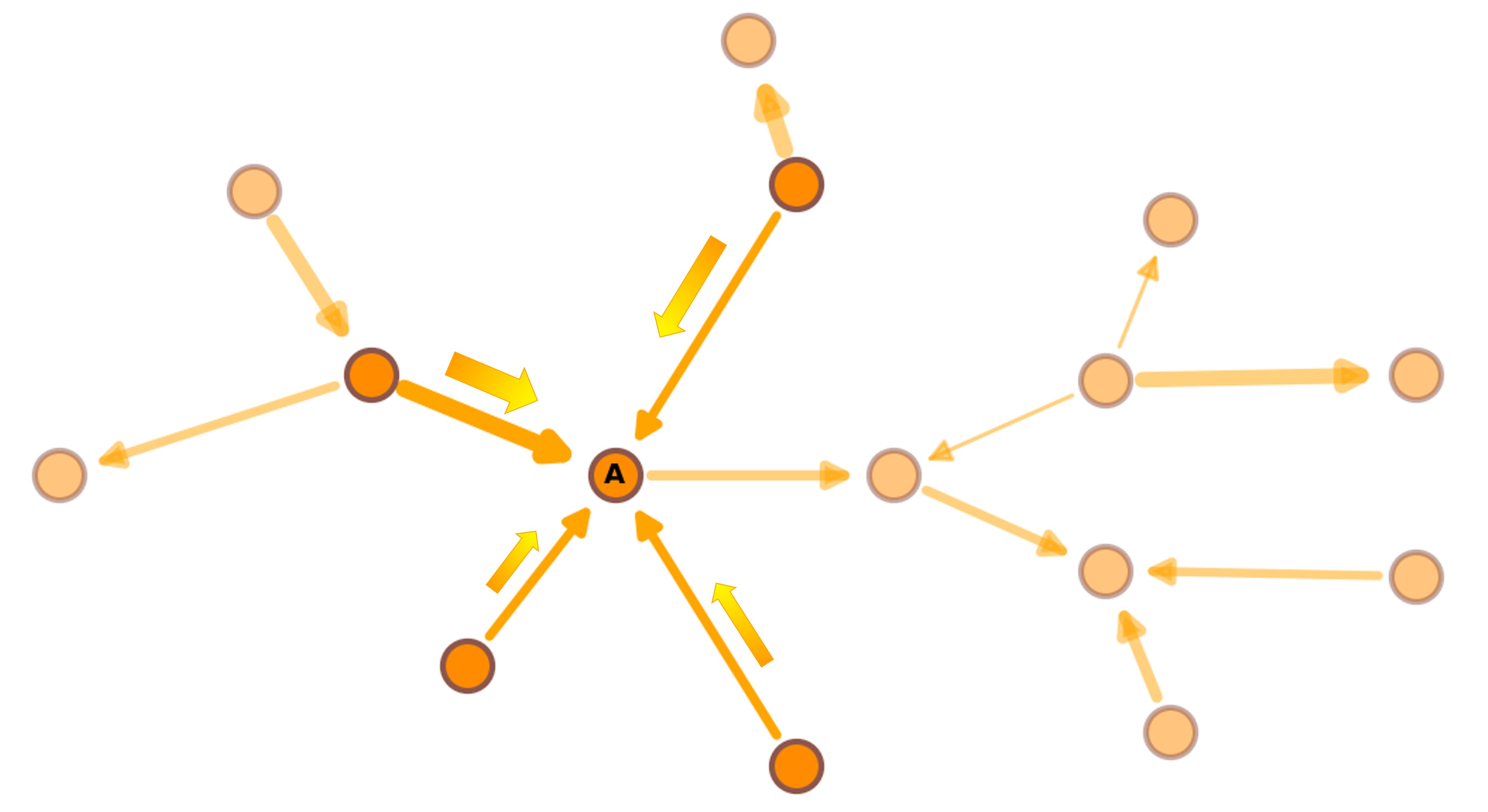}
    \caption{Risk Propagation Illustration for the \textit{Network Risk Estimation} system. Entity \textit{\textbf{A}}'s risk, which is an entry in risk vector $\mathbf{x}_t$, at the next time step is its previous risk plus the weighted sum of its neighbors' risks weighted by the edge weights given by functional connectivity graph $\mathbf{F}^{(t)}$. This risk propagation idea specifies the temporal and spatial dynamics of the risk.}
    \label{fig:risk_prop2} 
\end{figure}

This idea is summarized precisely as the set of linear equations given by (\ref{eq:risk_prop}), and it is illustrated in Figure \ref{fig:risk_prop2}.

\begin{equation}
    \mathbf{x}_{t + \tau} = \mathbf{F}^{(t)} \mathbf{x}_{t}
    \label{eq:risk_prop}
\end{equation}

Here $\mathbf{x}_{t}$ is the current risk estimates for the whole network at time $t$, $\mathbf{F}^{(t)}$ is the graph calculated from connections in time interval $[t, t+\tau)$ and $\mathbf{x}_{t + \tau}$ is predicted risks at time $t + \tau$. Another interpretation of (\ref{eq:risk_prop}) is a diffusion process on the weighted graph $\mathbf{F}^{(t)}$ where the risks are diffused through the edges in accordance with the edge weights. Ultimately, the risk propagation idea summarized by (\ref{eq:risk_prop}) states the temporal dynamics and spatial relationships of the entity risks, i.e., how the risks evolve in the network over time.

To give an example of the risk propagation model and the notion of entity risks, consider the threat of malicious-software/virus for a cyber network. The entity risks, in this case, can be the expected cost of a given entity being compromised by the virus in dollar amounts. Antivirus countermeasures are able to tell us about the likelihood of certain entities being compromised, and consequently the associated risk, but this is not the case for the whole network. Our intuition about how the virus advances among entities is essentially summarized by the saying \say{Tell me who your friends are, and I'll tell you who you are}, which constitutes the risk propagation model. Accordingly, if we have an estimate of the risk distribution for every entity in the network at some point in time, then we can have an educated guess about the risk distribution for the very next time step. This insight about the risks is not covered by the antivirus countermeasures, and it has to do with the observed behavior of entities.

\subsubsection{Optimal Linear Risk Estimator}
\label{sec:kalman}

Using the risk propagation model given in (\ref{eq:risk_prop}), one can keep predicting the risks given the connections and initial estimate. Even though this estimate carries the intuitive idea that entities influenced by risky entities should become riskier, the risk estimates are prone to errors. As an example, a small error in the initial risk estimate and an error in the estimation of the relationship graph will lead to increasing discrepancies between the actual risks and the risk estimates over time. For this reason, the risk estimates obtained by (\ref{eq:risk_prop}) are refined with the incoming risk measurements in minimum mean squared error sense, where the measurements are referred to collectively as $\mathbf{z}_t$. The resulting estimation process is also known as the discrete Kalman Filter \cite{kalman_book}. Integration of this probabilistic framework into the risk estimation problem is crucial since it not only gives the optimal risk estimates according to the assumed model but the uncertainties of those estimates as well, in the form of error covariance matrix $\mathbf{P}_{t|t}$.

Based on (\ref{eq:risk_prop}), a linear probabilistic model is assumed for the entity risks $\mathbf{x}_{t}$. A portion of the entities are risk-measured at time instance $t$, represented by $\mathbf{z}_{t}$.
\begin{equation} 
    \begin{split}
    \text{System and } & \text{Measurement Model:}\\
    \mathbf{x}_{t+\tau} &= \mathbf{F}^{(t)} \mathbf{x}_{t} + \mathbf{w}_t\\
    \mathbf{z}_t &= \mathbf{H}_t \mathbf{x}_{t} + \mathbf{v}_t\\
    \end{split} \label{eq:kalman_model}
\end{equation}

Here, $\mathbf{x}_t \in \mathbb{R}^n$ is the actual risks at the beginning of time $t$, $\mathbf{w}_t \sim \mathcal{N}(\mathbf{0}, \mathbf{Q}_t)$ is the system process noise which is independently normal distributed with $\mathbf{0}$ mean and diagonal covariance matrix $\mathbf{Q}_t$, $\mathbf{v}_t \sim \mathcal{N}(\mathbf{0}, \mathbf{R}_t)$ is the independently normal distributed measurement noise, $\mathbf{z}_t \in \mathbb{R}^m \;$ is the risk measurements and $\mathbf{H}_t \in \{0, 1\}^{m \times n}$ is the matrix that selects entities that are measured at time $t$. Essentially, each row of $\mathbf{H}_t$ and $\mathbf{z}_t$ corresponds to a risk measurement in the network. $\mathbf{z}_t$ contains the measured risk value, the respective row of $\mathbf{H}_t$ has a $1$ in the position of the measured entity, and the rest of the entries are $0$. $\mathbf{x}_t$ represents risk values of all entities in the network, which is the value desired to be estimated, whereas $\mathbf{z}_t$ are the noisy risk measurements from a usually small number of entities. In Figure \ref{fig:net}, $\mathbf{z}_t$ corresponds to risk measurements of red and green entities, and $\mathbf{x}_t$ represent the risks of every entity; red, green, gray and blue combined.

The estimate risk distribution is described by the mean of the estimate $\hat{\mathbf{x}}_{t|t} = \mathbb{E}[\mathbf{x}_t | \mathbf{z}_0^{t}]$ and the error covariance matrix $\mathbf{P}_{t|t} = \mathbb{E}\big[\big(\mathbf{x}_t - \hat{\mathbf{x}}_{t|t}\big)\big(\mathbf{x}_t - \hat{\mathbf{x}}_{t|t}\big)^T|\mathbf{z}_0^{t}\big]$ where $\mathbb{E[.]}$ is the expected value operator and $\mathbf{z}_0^{t} = \{\mathbf{z}_0, ..., \mathbf{z}_t\}$ are all the risk measurements up to time $t$. Assuming the model in (\ref{eq:kalman_model}) and an initial guess about the risk distribution, the linear optimal risk estimator is given recursively by (\ref{eq:kalman_predict}) and (\ref{eq:kalman_update}), which is derived in \ref{sec:app_kalman}. 

\begin{equation}
    \begin{split}
    \text{Predict:} &\\
    \mathbf{\hat{x}}_{t+\tau|t} &= \mathbf{F}^{(t)}\mathbf{\hat{x}}_{t|t}\\
    \mathbf{P}_{t+\tau|t} &= \mathbf{F}^{(t)} \mathbf{P}_{t|t}{\mathbf{F}^{(t)}}^T + \mathbf{Q}_t\\
    \end{split} \label{eq:kalman_predict}
\end{equation}

\begin{equation}
    \begin{split}    
    \text{Update:} &\\
    \mathbf{S}_{t+\tau} &= \mathbf{H}_{t+\tau}\mathbf{P}_{t+\tau|t}\mathbf{H}_{t+\tau}^T + \mathbf{R}_{t+\tau}\\
    \mathbf{K}_{t+\tau}^* &= \mathbf{P}_{t+\tau|t}\mathbf{H}_{t+\tau}^T\mathbf{S}_{t+\tau}^{-1}\\
    \mathbf{\hat{x}}_{t+\tau|t+\tau} &= \mathbf{\hat{x}}_{t+\tau|t} + \mathbf{K}_{t+\tau}^* (\mathbf{z}_{t+\tau} - \mathbf{H}_{t+\tau} \mathbf{\hat{x}}_{t+\tau|t})\\
    \mathbf{P}_{t+\tau|t+\tau} &= \mathbf{P}_{t+\tau|t} - \mathbf{K}^*_{t+\tau}\mathbf{S}_{t+\tau}{\mathbf{K}_{t+\tau}^*}^T
    \end{split} \label{eq:kalman_update}
\end{equation}

Risk distribution described by $\mathbf{\hat{x}}_{t|t}$ and $\mathbf{P}_{t|t}$ is the best linear risk estimate, in minimum mean squared error (MMSE) sense, in the presence of risk measurements \cite{kalman_optimal}. Initial risk estimate need not be exact and can be guessed, such as uninformative uniform mean $\mathbf{\hat{x}}_{0|0} = \mathbf{1}$ and process noise for error covariance $\mathbf{P}_{0|0} = \mathbf{Q}_{0}$. The Kalman Filter given by (\ref{eq:kalman_predict}) \& (\ref{eq:kalman_update}) guarantees that the estimation error will converge to zero in the case of fixed system parameters $\mathbf{F}, \mathbf{H}, \mathbf{Q}$ and $\mathbf{R}$ as long as they are accurate. If no measurement is made at a time step, then only the predict step is applied and \textit{a posteriori} estimate becomes the same as \textit{a priori} estimate, $\mathbf{\hat{x}}_{t+\tau|t+\tau} = \mathbf{\hat{x}}_{t+\tau|t}$ and $\mathbf{P}_{t+\tau|t+\tau} = \mathbf{P}_{t+\tau|t}$.

\subsubsection{ Relief Factor}
\label{sec:relief}

The risk estimates given by (\ref{eq:kalman_update}) tell us about the risk distribution in the network as a result of \textit{(i)} the risk propagation model and \textit{(ii)} risk measurements. Although the relative risks of entities are the more interesting aspect, without proper scaling, the risk estimates $\mathbf{\hat{x}}_{t|t}$ are unbounded. This happens when the functional connectivity graph $\mathbf{F}^{(t)}$ has eigenvalues greater than 1. From risk estimation point of view, this is an undesired effect since an entity left alone should not have unbounded risks when no additional measurements are provided. Consequently, this desired property is added to the estimation scheme by scaling the posterior estimates $\mathbf{\hat{x}}_{t+\tau|t+\tau}$ and $ \mathbf{P}_{t+\tau|t+\tau}$ with what's here referred as the \textit{relief factor} $\rho_{r}$ which acts as a design parameter of the risk estimation system.

\begin{equation}
\begin{split} 
    \mathbf{\hat{x}}_{t+\tau|t+\tau}' &= (1-\rho_{r})  \mathbf{\hat{x}}_{t+\tau|t+\tau}\\
    \mathbf{P}_{t+\tau|t+\tau}' &= (1-\rho_{r})^2 \mathbf{P}_{t+\tau|t+\tau}
\end{split} 
    \label{eq:relief_factor}
\end{equation}

The relief factor essentially controls the rate at which the risks of entities degrade when left unobserved. The estimates $\mathbf{\hat{x}}_{t+\tau|t+\tau}'$ and $\mathbf{P}_{t+\tau|t+\tau}'$ give the risk estimates accounted for the scaling effect of functional connectivity graph, and they can now be readily used for decision making concerning risks. As a rule of thumb, the value $\rho_{r}^* = 1-1/\lambda_{max}$ will theoretically ascertain that risks stay bounded where $\lambda_{max}$ is the largest eigenvalue of $\mathbf{F}^{(t)}$. The larger the value of $\rho_{r}$, the faster the risk estimates will diminish. The value of the relief factor can be adjusted just like any other hyper-parameter to meet the desired risk levels prior to testing. If no observations, $\mathbf{z}_{t}$, are provided, relative values of entity risks are conserved regardless of the value of the relief factor $\rho_{r}$.

\subsection{ Real Time Operation}
\label{sec:real_time_op}
A profound advantage of risk estimation over risk measurement is that estimation can be done as the network is observed, whereas measurements are much less frequent. In this section, the flowchart of the real-time operation of the proposed risk estimation system is outlined.

\begin{figure}[ht]
    \centering
    \includegraphics[width= \linewidth]{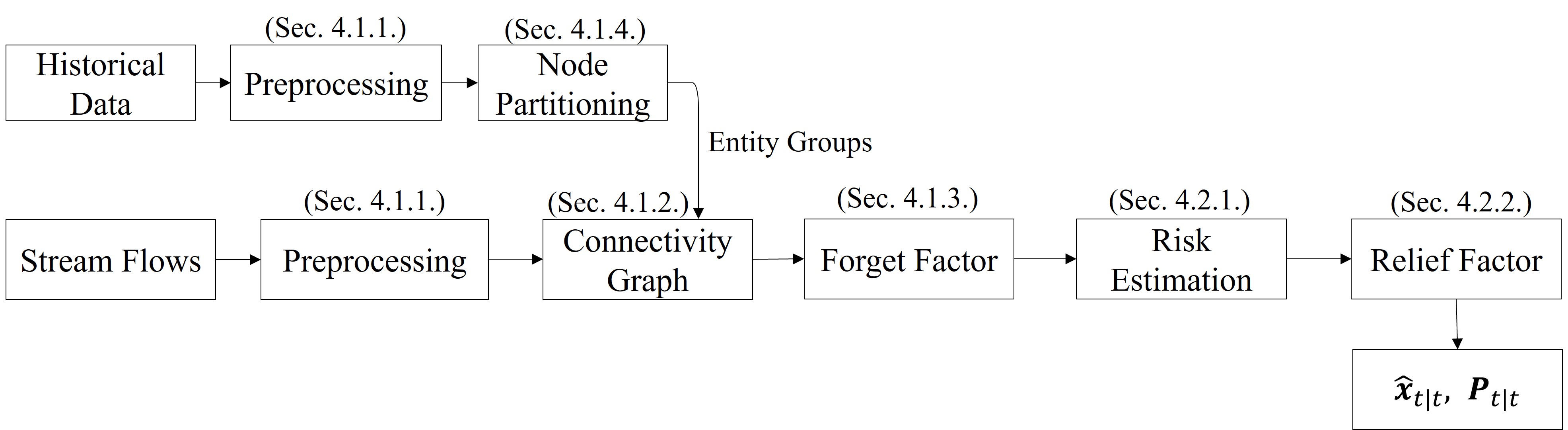}
    \caption{Flow chart for Real-Time Operation of \textit{NRE}. The top row is done offline on a large set of past flows to infer entity communities. The bottom row computes the functional connectivity graph for the current time window and estimates the risks as flows occur. Related section numbers are given above each block.}
    \label{fig:flow_chart}
\end{figure}

First, in order to reduce the possible computational overhead due to the size of the network, past representative flow data of the network is processed for the identification of related entity groups. The entity signals are formed as in Section \ref{sec:synch}, and then the functional connectivity graph for the whole network of interest is calculated as in Section \ref{sec:graph}. Then, the functional connectivity graph for the whole network is partitioned into non-overlapping groups of entities as described in Section \ref{sec:partition}, which defines the entity groups for the upcoming steps. This whole process is done offline as denoted by the top row of Figure \ref{fig:flow_chart}.

Once entity groups of desirable size are obtained, their flows are then processed similarly and current functional connectivity graphs for each cluster are calculated. Finally, \ul{for each entity cluster} risk prediction and update according to the provided risk measurements are employed as in Section \ref{sec:kalman} \ul{independently} and this latter process is illustrated as the bottom row of Figure \ref{fig:flow_chart}.

\section{Results and Discussion}
\label{sec:results}
The proposed risk estimation system has been experimented on a public dataset and is presented in this section. Most public datasets contain flow information and, occasionally, part of the network topology. However, risk measurements are problem-specific, and they are not commonly measured in gathering a dataset. The dataset used for testing is the CIC-IDS-2017 dataset \cite{cicids2017, cicids2017_dataset}, which emulates a real-life network with benign network traffic and multiple attacks over 5 days. The benign traffic of this dataset was generated by realistic user profiles, and the attacks included common intrusions that aim to exploit system vulnerabilities in order to gain leverage over the network for malicious purposes \cite{cicids2017}. This dataset has been used as the benchmark for testing new risk measurement methods \cite{chang2021graph, dias2020big, wu2022intrusion} .

Section \ref{sec:risk_ests} demonstrates the resulting risk estimates over time with synthetic risk measurements. In Section \ref{sec:application}, an example of possible network management action with the proposed risk estimates is illustrated. In Section \ref{sec:concept}, the effectiveness of the proposed risk estimates is compared to a risk measurement alternative, and finally, in Section \ref{sec:running}, a running time analysis of the proposed system is presented. 

\subsection{Risk Estimates Over Time}
\label{sec:risk_ests}
Monday Session of the CIC-IDS-2017 dataset contains benign flows produced by the inner network that consists of servers, firewall, routers, switches, and users  \cite{cicids2017}. This session is used for illustrating the risk estimation procedure since it resembles common network connection data, and artificial measurements based on the provided network topology were added to analyze the results of the proposed risk estimation method.

\begin{figure}
    \centering
    \includegraphics[width= \textwidth]{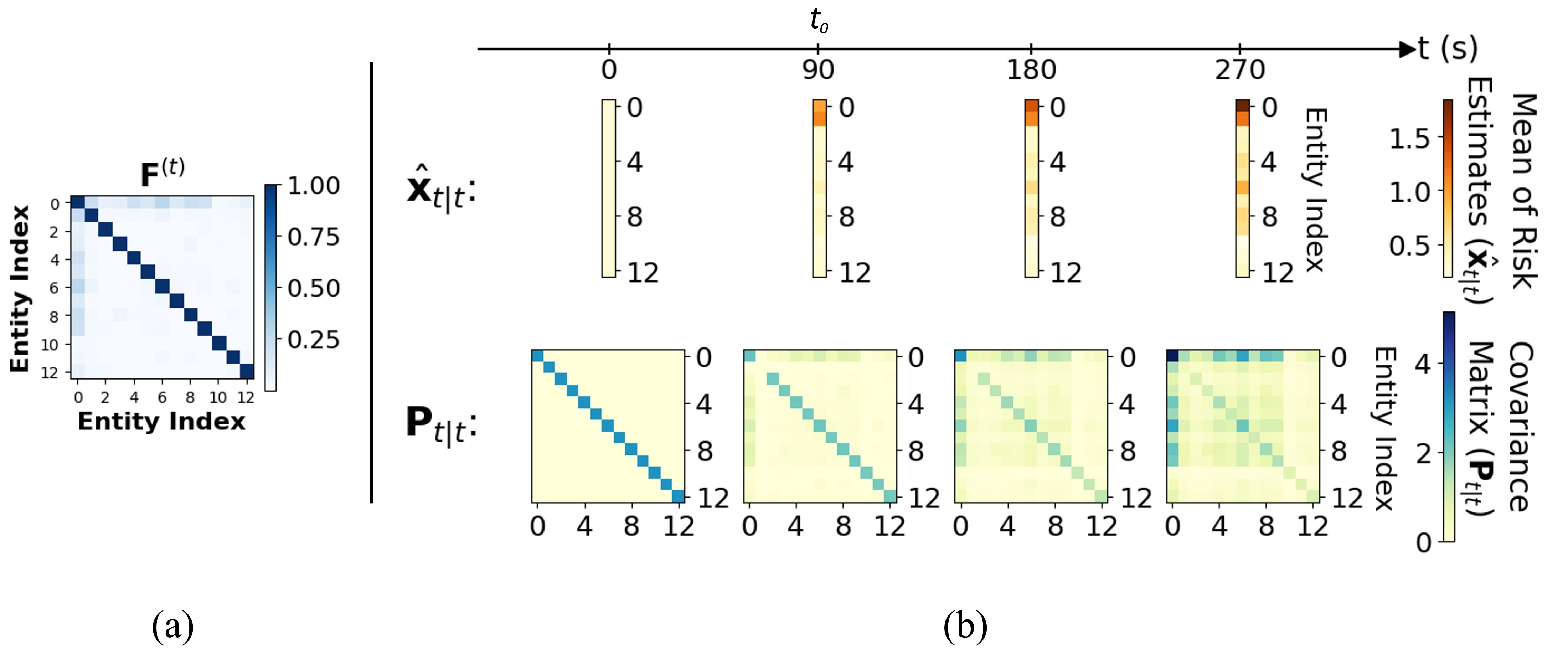}
    \caption{Risk Estimates over time for the \textit{Network Risk Estimation} system. (a) Functional Connectivity Graph $\mathbf{F}^{(t)}$ that is calculated from the flows of usual activity for this network and is held fixed during risk estimation. (b) Mean of risk estimates $\hat{\bm{x}}_{t|t}$ and covariance matrix $\bm{P}_{t|t}$ at the start of each synchronization time window. Low variance risk measurement is provided on entity one at time $t_0 = 90s$, which appears as the small value in the diagonal of the risk estimate covariance matrix $\bm{P}_{t|t}$. Over time, the mean of the risk estimate $\hat{\bm{x}}_{t|t}$ converges to a state dictated by the functional connectivity graph $\mathbf{F}^{(t)}$.}
    \label{fig:samp_res}
\end{figure}

In this part, the risk estimates for the inner network \cite{cicids2017} are calculated for a brief time interval under normal operation of this network. In this illustration, the functional connectivity graph $\mathbf{F}^{(t)}$ is computed from the flows that are comprised of benign emulated user activity, and it is held fixed during the risk estimation process for simplicity. Low uncertainty risk measurement of high-risk value is defined on the web server \textit{192.168.10.50}, indexed as entity 1, at time instance $t_0=90s$, and the relief factor is set slightly below the rule of thumb defined in Section \ref{sec:relief}. Snapshots of the calculated mean of relative risk estimates $\hat{\bm{x}}_{t|t}$ and their respective covariance matrices $\bm{P}_{t|t}$ are given in Figure \ref{fig:samp_res}b.

Initially, the relative risk estimates were set to independent identically normal distribution $\mathcal{N}(c\bm{1}, \sigma^2\bm{I})$ with low risk mean $c$ for each entity due to the lack of prior information about entity risks $\sigma^2$ being their variance. Single risk measurement was done at time instance $t_0=90s$ on entity 1 whose effect can be seen from \ref{fig:samp_res}b as high relative mean risk value for entity 1 and low variance, the (1,1) entry of risk the covariance matrix $\bm{P}_{t|t}$. After some time, the relative risk distribution converged to a stationary state given by the dominant eigenvector of the functional connectivity graph $\mathbf{F}^{(t)}$ since risk estimation without measurements simplifies to consecutive multiplication by the functional connectivity graph $\mathbf{F}^{(t)}$ that is also fixed. In this state, entity 0 with IP address \textit{192.168.10.3}, which turns out to be the DNS server in this network topology \cite{cicids2017}, ends up having the highest risk due to its higher connectivity with other entities as seen from the functional connectivity graph $\mathbf{F}^{(t)}$ on Figure \ref{fig:samp_res}a.

It is fair to say entity 0 has become the risk sink in this sub-network, which is along with our risk propagation intuition (\ref{eq:risk_prop}) since it is influenced by many entities and risk estimation is dominated by the prediction step (\ref{eq:kalman_predict}) as opposed to the measurements. The variances of risk estimates, the diagonal values of the covariance matrix $\bm{P}_{t|t}$, in general increase over time due to successive prediction, and the only mechanisms that reduce these variances are the introduction of risk measurements or the risk relief controlled by the relief factor where both were present in this example.

All in all, probabilistic risk distributions described by $\hat{\bm{x}}_{t|t}$ and  $\bm{P}_{t|t}$ carry all the information necessary to make decisions in this example since the prior risk distribution was specified as a Gaussian distribution. In the most general case, only the mean and covariance matrix of the prior risk distribution still needs to be provided to the proposed system, which can be chosen as a uniform mean with some variance as in this example if no information is available regarding the prior risk distribution. Either way, the risk estimates are optimal among all linear estimators given the risk propagation model (\ref{eq:kalman_model}).

As an example of utilization of these risk estimates in network management, $\hat{\bm{x}}_{t|t}$ can be used to spot the highest risk entity with the respective confidence given by $\bm{P}_{t|t}$. Alternatively, as illustrated in Section \ref{sec:application}, routing on this network can be designed to minimize over estimated risks.

\subsection{Simple Safe Routing}
\label{sec:application}
Another benefit of the proposed work \textit{NRE} is that since it is a quantitative risk assessment method, decision-making and management of the network is straightforward after obtaining the risk estimates. One example is the \say{safe routing} design, which is to pick the route for packets to traverse in the network when any two entities are set to communicate. This section provides a simple safe-routing application by finding the best route for a pair of entities based on a security measure of paths in the network topology and the proposed entity risk estimates.

\begin{figure}
    \centering
    \includegraphics[width= 0.75\linewidth]{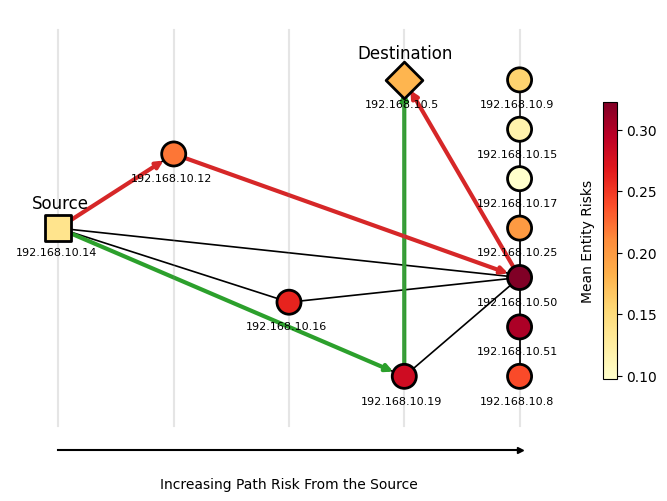}
    \caption{A solved case of simple safe routing for the insider network of CIC-IDS-2017 \cite{cicids2017}. Entities, with unique identifiers beneath, are sorted horizontally according to their min-max path's distance, called path risk. Node colors color-code the mean entity risks, and edges indicate the network topology or available paths. The green path is the min-max risk path between the source entity and the destination, and the red path is not since \textit{192.168.10.50}'s risk is higher than every other entity in the green path.}
    \label{fig:safe_routed} 
\end{figure}

It is assumed that the underlying network topology is known; that is, for each entity, the entities that can be reached are given. Let the mean risk estimates at the current time be estimated. Then, intuitively, a user in the network will desire to avoid high-risk entities and prefer low-risk entities to route the desired packets over, as illustrated in Figure \ref{fig:net}. Given the risk estimates, this objective can be formalized as minimizing a security loss over the available paths in the network topology, referred to as the \textit{path risk}. Different security losses such as the \textit{average mean risk of entities along the path} are reasonable, and the one that is chosen in this example is \textit{maximum mean risk of entities along the path}. For this choice, the packets will be routed with the set of entities in the network for which all the mean risks are below a certain level, which is made as small as possible by the path choice. This path will be referred to as the \textit{min-max path}. For the sub-network in Section \ref{sec:risk_ests}, excluding the DNS server, the min-max path between a source and every other entity is computed, and the result is depicted in Figure \ref{fig:safe_routed}.

Here, the leftmost entity \textit{192.168.10.14} is the source from which all min-max paths are calculated. Every other entity is arranged according to their min-max path's distance from the source, the path risks and their identifier IP addresses beneath them. If the packets are desired to be sent to the entity \textit{192.168.10.5}, then the green path in Figure \ref{fig:safe_routed} is the path with the least maximum risk, which is the choice for simple safe routing problem.

In the identification of safe paths, only the mean of the risk estimates $\hat{\bm{x}}_{t|t}$ was used, and thus it was called the \say{simple safe routing}. In a more general setting, the risk distribution described by the error covariance matrix $\bm{P}_{t|t}$ can be utilized, and also, the objective loss that is to be minimized can be designed as a combination of path risk and round trip time signifying a trade-off between security and packet delay. 

Ultimately, this example illustrates a network management application of the proposed risk estimates, highlighting a way to add \say{security} aspect to a network function. With this addition, network decisions can be judged by previously unknown risks, and utility vs. security allotment can be quantitatively assessed. In the case of routing, in some cases, it might be desirable to sacrifice slightly from the round trip delay to keep a path's risk level lower. Having risk estimates described by $\hat{\bm{x}}_{t|t}$ and $\bm{P}_{t|t}$ enables this design.

\subsection{ Proof of Concept} 
\label{sec:concept}
The method outlined in Section \ref{sec:method} can be used to estimate the risks of individual entities that are not necessarily risk-measurable but are observable in terms of the connection data. A fair question is whether these risk estimates convey substantial information about the network state when the risk measurements are limited. In this section, the utility of these estimates is assessed by an experiment designed on the same labeled dataset CIC-IDS-2017 \cite{cicids2017_dataset}. In particular, \textit{Tuesday} session has been selected for this experiment since it contained the most different variety of threats.

The experiment is a binary classification problem of identifying the state of the network, whether malicious activity is present or not, labeled as \textit{\say{ATTACK}} or \textit{\say{BENIGN}} accordingly. The risk estimates obtained by the proposed method on the identified relevant part of the network have been compared to the limited risk measurements from a trained model on the flows of the insider network \cite{cicids2017} for this binary classification problem.

\subsubsection{Experimental Setup}
For this experiment, the dataset is divided into chunks of flows consisting of consecutive flows in designated time windows for calculating the functional connectivity graph, $\mathbf{F}^{(t)}$, and the state of the network is assigned either as \textit{\say{ATTACK}} if any of the flows are part of an attack or as \textit{\say{BENIGN}} otherwise. The proposed risk estimation method is then employed to estimate the risks in the network without any measurements provided. The network is partitioned according to the entity behaviors in the attack-free \textit{Monday} session flows of the dataset. Specifically, the node partitioning, as described in Section \ref{sec:partition}, is performed on these flows, which led to the discovery of 141 entities relevant to the insider network. These entities are then set as the sub-network for the risk estimation process on the \textit{Tuesday} session.

\begin{figure}[ht]
    \centering
    \includegraphics[width= \linewidth]{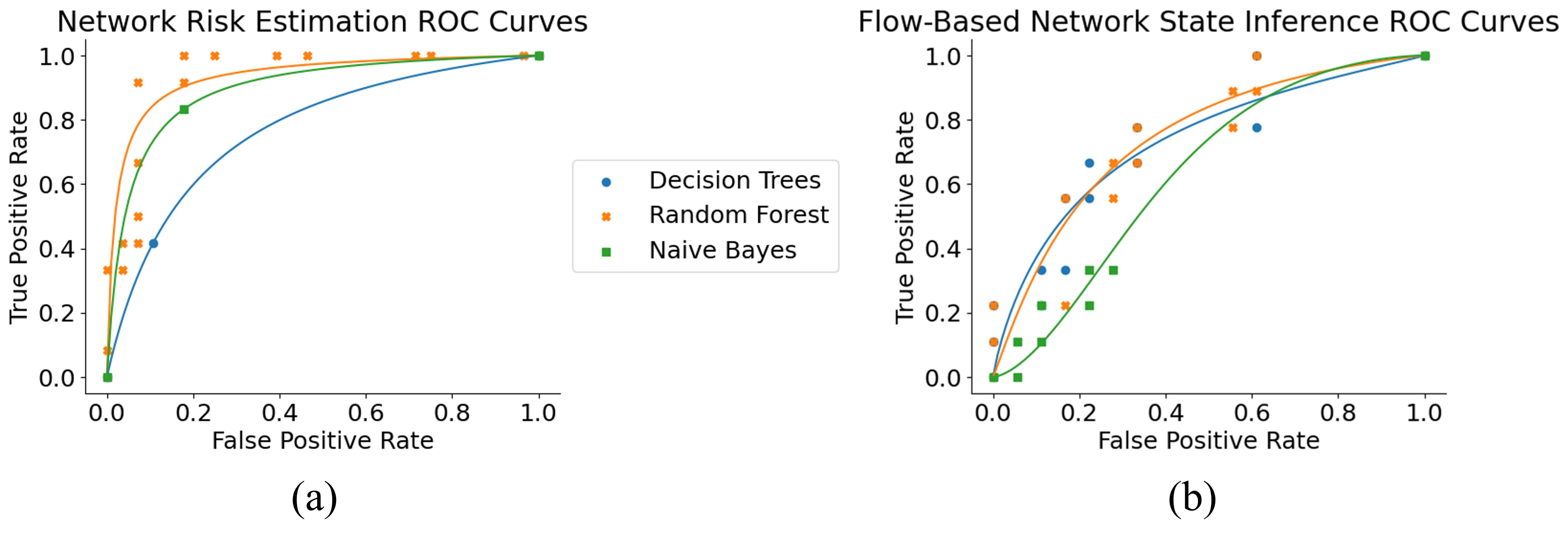}
    \caption{Comparison of ROC curves on validation set for the \protect\say{\textit{Number of Packets Received}} connection parameter. (a) Proposed Method \textit{NRE}: Network state can be inferred with as good as $0.95$ AUC (Area Under the Curve) when risks are estimated for the related part of the network. (b) Risk Measurement Method \textit{FBNSI}, based on \cite{li2020cloud, wu2022intrusion, tvmm}: Network state inference is possible for the measurable part of the network but with lower performance with the best AUC being $0.77$.}
    \label{fig:rocs} 
\end{figure}

The chunks of \textit{Tuesday} session are split into training, validation, and test sets for which the mean of entity-level risks are calculated for a particular connection parameter, as detailed in Section \ref{sec:method}, without any measurements provided. Then the risk estimate vector $\mathbf{\hat{x}}_{t|t} \in \mathbb{R}^{141}$ is used as a sample in 141-dimensional feature space to train binary classifiers using three different generic machine learning models of various complexity: Naive Bayes, Decision Trees, and Random Forest \cite{scikit-learn}. The experimental \say{Receiver Operating Characteristics} (ROC) curves of these classifiers are then obtained on the validation set, and the hyperparameters (\textit{i}) synchronization window size $\delta$, (\textit{ii}) the window size $\tau$ for graph calculation, and (\textit{iii}) the forget factor $\rho_f$ are tuned via maximizing the \say{area under the curve (AUC)} performance metric, for the best classifier among the aforementioned models, on the validation set. The relief factor is not considered in tuning for this experiment since it has no effect on classification performance but rather is just a scaling factor of the features due to the absence of risk measurements. After a grid search on the space of potential hyper-parameters, the combination $(\delta,\: \tau,\: \rho_f ) = (1.2s,\: 90s,\: 0.5)$ is picked. For these hyperparameters, an instance of ROC curves obtained for the proposed method (\textit{NRE}) with \say{Number of Packets Received} connection parameter is given in Figure \ref{fig:rocs}a. The tuned hyper-parameters are then used on the previously separated test session for the goodness assessment of the proposed risk estimates in network state inference. 

The proposed method is compared to a representative of previous works in risk measurement, which was restricted to measurable entities. The previous work on network state inference considers various flow attributes and builds a model from these attributes for estimating the risk of the network as a whole \cite{li2020cloud, wu2022intrusion, tvmm}. In this experiment, this method is referred to as the \say{Flow-Based Network State Inference} method (\textit{FBNSI}) due to its similarity with the flow-based risk measurement methods in Intrusion Detection literature \cite{sperotto2010overview}. To capture the limitations of the sole risk measurement methods in practice, flows of the \textit{FBNSI} are constrained to the flows of the insider network in this experiment. This sub-network contained only 13 entities, including the DNS server \cite{cicids2017}. The insider network is the local network that was the target of the emulated attacks in forming this dataset \cite{cicids2017}. 

Previous work in risk measurement would extract information such as average round trip delay, transmission rate, and response time \cite{li2020cloud} from flows to predict the risk of the whole network via predicting the likelihood of an attack. In contrast, the proposed system is designed to make use of per-entity attributes to estimate entity-level risks, which then tell about the network state. Consequently, in this experiment, the risk measurement method \textit{FBNSI} is designed to make inferences, \textit{BENIGN} or \textit{ATTACK}, about the state of the network using the same time windows as the proposed method. In order to keep the information used about the flows same, the flow attributes are restricted to the same connection parameter used by the proposed method, e.g., the number of packets in forward and backward direction flow attributes are selected for the \say{Number of Packets Received} connection parameter. As a result, the feature space for the risk measurement method \textit{FBNSI} is either one or two-dimensional, whose features correspond to the flow attribute used to generate the risk estimates by the proposed method.

\begin{figure}[ht]
    \centering    
    \includegraphics[width= 0.9\linewidth]{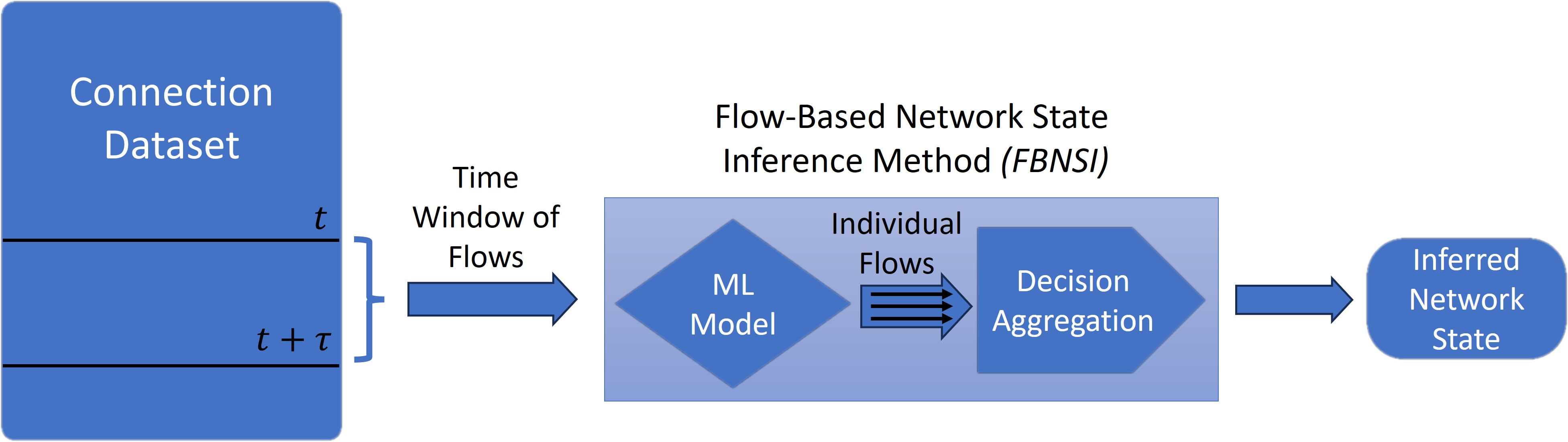} 
    \caption{Flowchart of the Flow-Based Network State Inference (\textit{FBNSI}) method. Chunks of flows for the time window $[t,t+\tau)$ are fed into a Machine Learning model trained on individual flows. The decisions on the individual flows are aggregated by counting the percentage of flows labeled as \textit{attack flow}, which determines the network state for the time window.}
    \label{fig:flow_based_chart} 
\end{figure}

Decisions are made about the network state with \textit{FBNSI} method by first training the same classifiers on individual flows, which classify the individual flows themselves as either \textit{benign flow} or \textit{attack flow}. For the time window of flows over which the network state is sought, the flow-level decisions are aggregated by simply counting the \textit{attack flow} percentage in the chunk. This percentage is assigned as the likelihood of the network state being \textit{ATTACK}, which is used in inferring the state of the network. The whole network state inference process is summarized in Figure \ref{fig:flow_based_chart}.

The only hyperparameter of this method is the time window size $\tau$ for network state estimation, which is picked as $\tau = 180s$ for testing after a similar tuning process on the training and validation set. The experimental ROC curves for the \textit{FBNSI} method are obtained via thresholding the likelihood of network state on the validation set, which is shown in Figure \ref{fig:rocs}b for \say{Number of Packets Received} connection parameter. For testing, the operating points for both methods are picked from the inferred curves as the points maximizing \textit{balanced accuracy} classification metric, which corresponds to the $45^{\circ}$ tangent to the ROC curves. Balanced accuracy has been chosen as the main classification performance metric since it accounts for the uneven distribution of classes, i.e., network states.

Finally, both methods are tested on the unseen test set from the \textit{Thursday} session of the dataset, and only the peak balanced accuracy among the three classifiers used is reported for clarity. The threats for this day consisted of web attacks such as \textit{Brute Force}, \textit{XSS}, and \textit{Sql Injection} \cite{cicids2017}. The testing was done using all available connection parameters except \say{Activation} since it is not available for \textit{FBNSI}. The window size $\tau$ was picked as $\tau=180s$ in order to compare two methods which was the optimal window size for \textit{FBNSI} but sub-optimal for \textit{NRE}. Figure \ref{fig:conn_sweep} shows these peak balanced accuracy across different connection parameters, which reflect how the model would do if both methods modeled different aspects of flows.

\subsubsection{Analysis}

The scatter points in Figure \ref{fig:rocs} are the experimental operating points yielded by thresholding the likelihood of the positive class given by the two methods described. For visual interpretation, smooth non-linear curves have been fitted to the operating points obtained on the validation set by fitting a line to the \textit{logit} transform of both coordinates. Inferred ROC curves show how these models will perform if more data from the same distribution were tested.

\begin{figure}[ht]
    \centering
    \includegraphics[width= 0.7\linewidth]{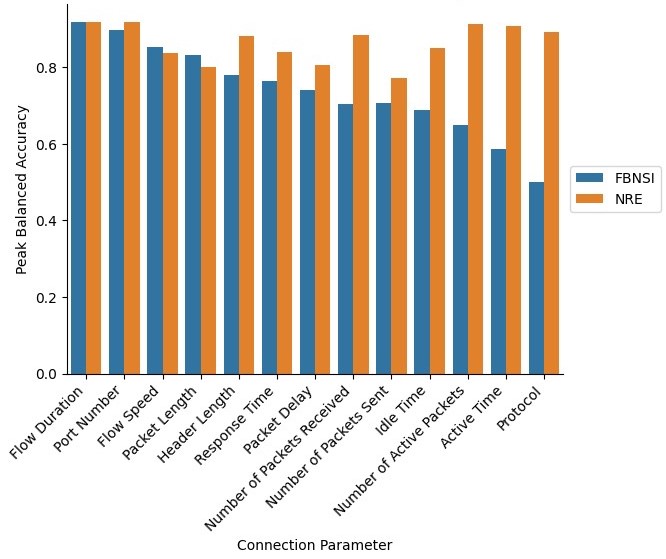}
    \caption{Comparison of purely risk measurement approach \textit{FBNSI} and proposed risk estimation method \textit{NRE} across different connection parameters in terms of peak balanced accuracy. When risk measurements are descriptive of the network (connection parameters towards the left), the risk measurement approach is viable. However, when risk measurements are limited, \textit{FBNSI} is not descriptive of the network. The proposed risk estimation approach \textit{NRE} is descriptive of network state regardless of the modeled aspect of flows with comparable network state inference performance.}
    \label{fig:conn_sweep} 
\end{figure}

For the test results, it is concluded from Figure \ref{fig:conn_sweep} that not all aspects of connection data had substantial information about the network state for the risk measurement method \textit{FBNSI}. As evident from the nature of the threats present in this session, some connection parameters, such as \say{Flow Duration} and \say{Port Number}, were sufficient in distinguishing the network state. However, connection parameters such as \say{Active Time} were not informative. In other words, different connection parameters in this experiment represent varying degrees of risk measurement expressiveness, and the performance of the risk measurement paradigm depends on the amount of information the measurements have about the network state. In the case of limited risk measurements, the model does not provide valuable information about the network state.

However, with the addition of entity identity and the propagation of the risk estimates to the rest of the network based on entity association, even the poor aspects of connection data were sufficient in inferring the network state, which is indicated as the consistently high peak balanced accuracy for \textit{NRE} in Figure \ref{fig:conn_sweep}. In other words, judging the network state by its entity's state and modeling their interactions as risk propagation have enhanced the visibility of the network in the case of poor risk measurements. 

The experiment conducted on the CIC-IDS-2017 dataset indicated that the proposed entity risk estimation method is able to provide sufficient information needed for network state inference. Estimating the entity risks has made detecting abnormal network states easier, even if the available risk measurements were not substantial. The proposed risk estimates maintain the crucial information needed to make judgments about the network, such as intrusion detection, while carrying all the benefits of being a quantitative risk assessment tool so that it can readily be used in network management and decision-making. 

\subsection{ Running Time Tests}
\label{sec:running}
One of the main advantages of risk estimation over risk measurement is operating in time scales that the device network might change. For this reason, it is desirable for the risk estimation system to have low running time complexity. Accordingly, the running time of the proposed risk estimation method has been analyzed over various design parameters on the same \textit{Tuesday} session of the CIC-IDS-2017 dataset \cite{cicids2017_dataset}. In this analysis, only the running time of calculation of the functional connectivity graph $\mathbf{F}^{(t)}$ using connection data is examined since the predict and update steps given by the Section \ref{sec:kalman} are just matrix multiplications that can be computed much faster than the aforementioned step. The experiments were conducted on Google Colaboratory environment \cite{bisong2019google} with CPU specifications being \textit{dual-core Intel(R) Xeon(R) CPU @ 2.20GHz.}

\begin{figure}[ht]
    \centering
    \begin{minipage}{.44\linewidth}
    \centering
    \includegraphics[width=\linewidth]{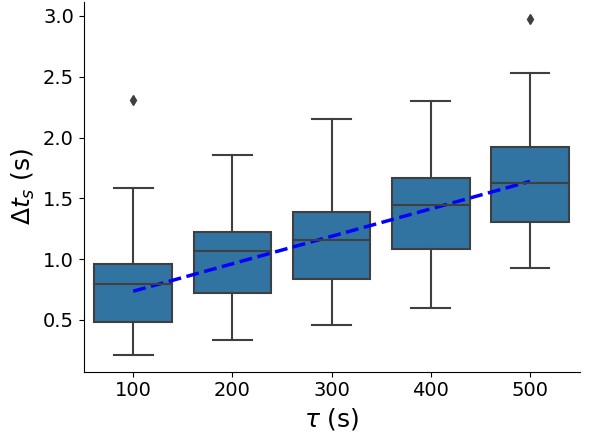}
    \textit{(a)}
    \end{minipage}
    \begin{minipage}{.452\linewidth}
    \centering
    \includegraphics[width= \linewidth]{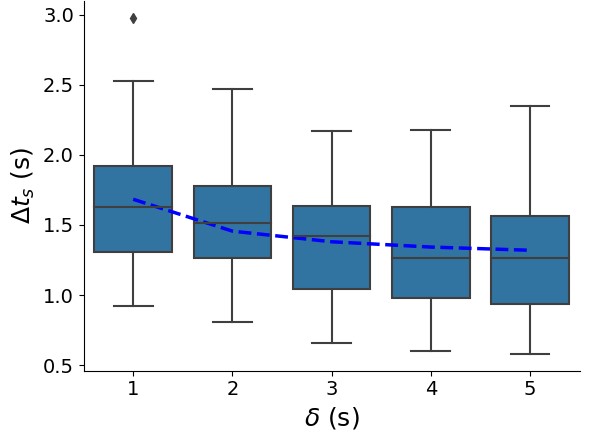}
    (\textit{b})
    \end{minipage}
    \caption{\protect\textit{(a)} Simulation Time vs. Graph Window Size with $1s$ Synchronization Windows. Running is proportional to the window size that is used to calculate the functional connectivity graph. \protect\textit{(b)} Simulation time vs. Synchronization Window size with $500s$ Graph Time Windows. Running time is inversely proportional to the window size that is used to aggregate flows to form the synchronous signals. For both cases, the simulation time $\Delta t_{s} \propto N$ where $N = \lfloor \tau/\delta \rfloor$ and the dotted line highlights this trend of the mean.}
    \label{fig:t_v_windows} 
\end{figure}

First, the running time per connection window was tested with respect to both time windows, the signal synchronization window size $\delta$, and graph window size $\tau$ as defined in Section \ref{sec:relationship_inference}. If the number of samples used to calculate a single edge weight in this graph is denoted as $N$, then it follows that $N=\lfloor \tau/\delta \rfloor$. A simple analysis of (\ref{eq:F_cov}) shows that the running time complexity of the calculation of a single edge in the connectivity graph is $\mathcal{O}(N)$ for the Pearson correlation coefficient estimator, although it might be different for others estimators used to calculate the functional connectivity graph $\mathbf{F}^{(t)}$. Figure \ref{fig:t_v_windows} shows the simulation time it took to calculate a single connectivity graph $\mathbf{F}^{(t)}$ for different time window sizes where the simulated running times are the box-plot, and the expected run-time is the dashed line.

\begin{figure}[ht]
    \centering
    \begin{minipage}{.45\linewidth}
    \centering
    \includegraphics[width=\linewidth]{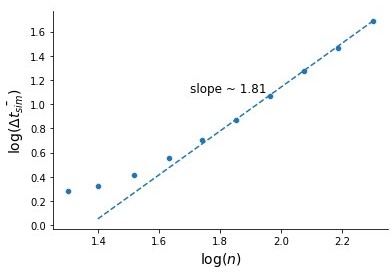}
    (a)
    \end{minipage}    
    \begin{minipage}{.44\linewidth}
    \centering
    \includegraphics[width=\linewidth]{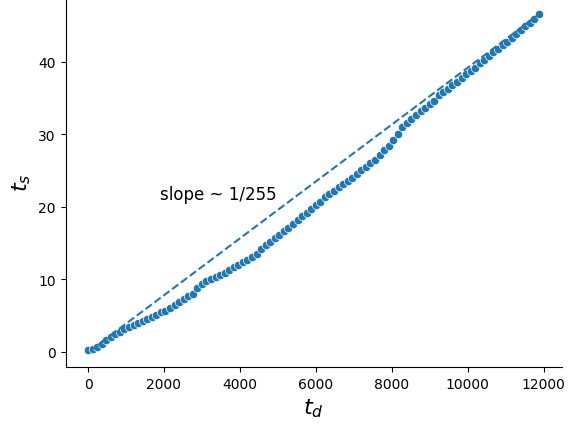}
    (b)
    \end{minipage}
    \caption{\textit{(a) }$\log$ Average running time vs. $\log$ Number of Entities in the network of interest. A dashed line with slope $1.81$ is the line fitted to the tail of the running times obtained, which implies an experimental bound on running time of $\mathcal{O}(n^{1.81})$. \textit{(b)} Collection of time instances that mark the start of functional connectivity graph estimation in the dataset, $t_{d}$, and the respective recorded simulation times, $t_{s}$. The average slope of $1/255$ indicates that $1$ seconds of simulation time is able to cover $255$ seconds of connection data, which is an indicator of real-time operability.}
    \label{fig:t_n_t_sim} 
\end{figure}

The other parameter that crucially affects running time is the size of the network or the number of entities in the connectivity graph $n$. In calculating the graph $\mathbf{F}^{(t)}$, each edge weight calculation is done separately, so the theoretical asymptotic behavior of the running time with respect to the size of the network is $\mathcal{O}(n^2)$ where the experiment showed even a tighter bound of $ \mathcal{O}(n^{1.81})$ given by Figure \ref{fig:t_n_t_sim}a.

Finally, for a particular parameter collection $(\tau, \delta, n) = (500s, 5s, 17)$ collection of timestamps at the start of the graph window in the dataset $t_{d}$ and the simulation times $t_{s}$ has been collected and is given in Figure \ref{fig:t_n_t_sim}b. This plot has an average slope of about $1/255$, which indicates that 1 second of simulation time covers 255 seconds of flows in the dataset. The gain of 255 implies that real-time operation is possible for this parameter set. 

For the real-time operation, it should be noted that parameter selection will be the deterministic factor for this gain. For instance, $\mathcal{O}(n^2)$ behavior in entity size implies that there will be a critical entity size $n^*$, with a simulation time to recording time gain of 1, beyond which no real-time operation will be possible. This scaling issue can be controlled by limiting the sub-network sizes in the node partitioning step in Section \ref{sec:partition}.

\section{Conclusion}
The risk measurement approach in cyber-network security is subject to loss in efficacy when the measurements are limited to a small subset of entities in the network and are coarse in time, both of which are common in practice. In this work, a risk estimation paradigm for cyber network security, NRE, is proposed as an alternative to existing risk measurement strategies. The proposed system NRE estimates the risks of entities based on their observed relationships in the connection data. For the case where risk measurements are available on some entities, the estimated risks are refined by the probabilistic framework using the optimal linear estimator. The end result of this system is the estimated risk distribution of the entities in terms of the mean estimate $\hat{\bm{x}}_{t|t}$ and the respective error covariance matrix $\bm{P}_{t|t}$ which carry all the information needed for active network management. The proposed system is fully data-driven as it does not need continuous human intervention, and it is adaptive to dynamic network environments where entity behavior might change over time.

The risk estimation process has been carried out on a public dataset. A direct network security management application of the proposed risk estimates called \say{simple safe routing} is illustrated on the same network topology. The proof of concept has been established on the same dataset, where it has been shown that, if needed, abnormal network states can be better inferred with the proposed method when the available risk measurements are limited. Finally, the real-time applicability of the proposed method was tested, which suggested that the online operation is possible with an appropriate choice of design parameters.

\appendix
\section{Derivation of Optimal Linear Risk Estimator}
\label{sec:app_kalman}
In this section, a derivation of the optimal risk estimator is presented, assuming the risk propagation model given in (\ref{eq:risk_prop}). The result is the discrete Kalman Filter applied for the risk estimation problem. The system and measurement model is given in (\ref{eq:kalman_model}), which is based on (\ref{eq:risk_prop}) with the process and measurement noise modeled. The goal is to find the unbiased linear estimator with minimum mean square error (MMSE) for the model in (\ref{eq:kalman_model}) given the risk measurements up to the current time $t$. This is also the optimal estimator when risks $\mathbf{x}_t$ are Gaussian.

Let $\hat{\mathbf{x}}_{t|t}$ be the best-unbiased estimate of the risks at time $t$ in the MMSE sense. It can be easily shown that this best estimate, in the MMSE sense, is the conditional expectation of the risks conditioned on the measurements up to time $t$ \cite{kalman_optimal}: 
\begin{equation}
    \hat{\mathbf{x}}_{t|t} = \mathbb{E}[\mathbf{x}_t | \mathbf{z}_0^{t}].
\end{equation} 
Here, $\mathbf{z}_0^{t} = \{\mathbf{z}_0, ..., \mathbf{z}_t\}$ is the collection of discrete risk measurements starting from time $0$, which is the reference point in time when the first measurement was made. The risk estimate  $\hat{\mathbf{x}}_{t|t}$ then has the error covariance matrix $\mathbf{P}_{t|t}$:
\begin{equation}
    \begin{split}
        \mathbf{P}_{t|t} %&= \mathbb{E}\big[\big(\mathbf{x}_k - \mathbb{E}[\mathbf{x}_k]\big)\big(\mathbf{x}_k - \mathbb{E}[\mathbf{x}_k]\big)^T|\mathbf{z}_1^{k}\big]\\
        &= \mathbb{E}\big[\big(\mathbf{x}_t - \hat{\mathbf{x}}_{t|t}\big)\big(\mathbf{x}_t - \hat{\mathbf{x}}_{t|t}\big)^T|\mathbf{z}_0^{t}\big]
    \end{split}    
\end{equation}
Let $\hat{\mathbf{x}}_{t+\tau|t}$ be the best estimator after a time window of length $\tau$, given the measurements up to time $t$. Then similarly using system model in (\ref{eq:kalman_model}) it can be shown that \cite{kalman_book},
\begin{equation}
\begin{split}
    \hat{\mathbf{x}}_{t+\tau|t} %&= E[\mathbf{x}_{t+\tau} | \mathbf{z}_0^{t}]\\
    %&= E[\mathbf{F}_{k}\mathbf{x}_{k} + \mathbf{w}_{k}| \mathbf{z}_1^{k}]\\
    %&=  \mathbf{F}_{k}E[\mathbf{x}_{k}|\mathbf{z}_1^{k}] + E[\mathbf{w}_{k}| \mathbf{z}_1^{k}]\\
    &= \mathbf{F}^{(t)} \hat{\mathbf{x}}_{t|t}.
\end{split} \label{eq:prior_mean}
\end{equation} 
This estimator in (\ref{eq:prior_mean}) is also known as the \textit{a priori} estimate since the current measurement $\mathbf{z}_{t+\tau}$ has not been accounted for yet. Similarly, the prior estimate covariance matrix can be calculated as \cite{kalman_book}:
\begin{equation}
\begin{split}
    \mathbf{P}_{t+\tau|t} %&= E\big[\big(\mathbf{x}_{k+1} - \hat{\mathbf{x}}_{k+1|k}\big)\big(\mathbf{x}_{k+1} - \hat{\mathbf{x}}_{k+1|k}\big)^T|\mathbf{z}_1^{k}\big]\\
    %&= E\big[\big( \mathbf{F}_{k}(\mathbf{x}_{k} - \hat{\mathbf{x}}_{k|k}) + \mathbf{w}_k\big)\big(\mathbf{F}_{k}(\mathbf{x}_{k} - \hat{\mathbf{x}}_{k|k}) + \mathbf{w}_k\big)^T|\mathbf{z}_1^{k}\big]\\
    %&= \mathbf{F}_{k} E\big[\big(\mathbf{x}_k - \hat{\mathbf{x}}_{k|k}\big)\big(\mathbf{x}_k - \hat{\mathbf{x}}_{k|k}\big)^T|\mathbf{z}_1^{k}\big] \mathbf{F}_{k}^T + E[\mathbf{w}_k\mathbf{w}_k^T|\mathbf{z}_1^{k}]\\
    &= \mathbf{F}^{(t)} \mathbf{P}_{t|t} {\mathbf{F}^{(t)}}^T + \mathbf{Q}_{t}.
\end{split} \label{eq:prior_cov}
\end{equation}

Equations (\ref{eq:prior_mean}) and (\ref{eq:prior_cov}) are known as the predict steps. In accounting for the effect of new observations about current risks, an \textit{unbiased} linear estimator is sought, which can be shown to have the following form \cite{kalman_optimal}:

\begin{equation}
\begin{split}
    \hat{\mathbf{x}}_{t+\tau|t+\tau} &= \hat{\mathbf{x}}_{t+\tau|t} + \mathbf{K}_{t+\tau}\mathbf{\nu}_{t+\tau}\\
    \mathbf{\nu}_{t+\tau} &= \mathbf{z}_{t+\tau}-\mathbf{H}_{t+\tau}\hat{\mathbf{x}}_{t+\tau|t}
    \end{split}
    \label{eq:post_mean}
\end{equation}
Here, the variable $\mathbf{\nu}_{t+\tau}$ is referred to as \textit{innovation} since it represents the difference between actual measurement $\mathbf{z}_{t+\tau}$ and the predicted measurement $\mathbf{H}_{t+\tau}\hat{\mathbf{x}}_{t+\tau|t}$ based on the knowledge up to time step $t$. $\mathbf{K}_{t+\tau}$ is the gain matrix for innovation which controls the contribution of measurements versus the previous estimate. (\ref{eq:post_mean}) describes a set of linear estimators. As the final step, the value of the gain matrix is specified by the MMSE criteria:

\begin{equation}
    \begin{split}
        \mathbf{K}_{t+\tau}^* &= \arg \min_{\mathbf{K}_{t+\tau}} \mathbb{E}[(\mathbf{x}_{t+\tau} - \mathbf{x}_{t+\tau|t+\tau})^T(\mathbf{x}_{t+\tau} - \mathbf{x}_{t+\tau|t+\tau})|\mathbf{z}_0^{t+\tau}]\\
        %&= \arg \min_{\mathbf{K}_{k+1}} trace(\mathbf{P}_{k+1|k+1})
    \end{split}
\end{equation}
Minimization has a closed-form solution which is known as the \textit{Kalman Gain} \cite{kalman_book}:
\begin{equation}
    \begin{split}
        \mathbf{K}_{t+\tau}^* &= \mathbf{P}_{t+\tau|t}\mathbf{H}_{t+\tau}^T\mathbf{S}_{t+\tau}^{-1}\\
        \mathbf{S}_{t+\tau} &= \mathbf{H}_{t+\tau}\mathbf{P}_{t+\tau|t}\mathbf{H}_{t+\tau}^T + \mathbf{R}_{t+\tau}
    \end{split}\label{eq:K_gain}
\end{equation}
The matrix $\mathbf{S}_{t+\tau}$ in (\ref{eq:K_gain}) is the \textit{innovation variance} and it can be shown that $\mathbf{S}_{t+\tau} = \mathbb{E}[\mathbf{\nu}_{t+\tau} \mathbf{\nu}_{t+\tau}^T]$. Combining with (\ref{eq:post_mean}) the error covariance matrix  $\mathbf{P}_{t+\tau|t+\tau} $ of the \textit{a posteriori} estimate $\hat{\mathbf{x}}_{t+\tau|t+\tau}$ can be obtained \cite{kalman_book}:

\begin{equation}
    \begin{split}
        \mathbf{P}_{t+\tau|t+\tau}
        &= \mathbf{P}_{t+\tau|t} - \mathbf{K}^*_{t+\tau}\mathbf{S}_{t+\tau}{\mathbf{K}_{t+\tau}^*}^T
    \end{split}\label{eq:post_cov}
\end{equation}

Together (\ref{eq:post_mean}), (\ref{eq:K_gain}) and (\ref{eq:post_cov}) are the update equations for the risk estimates. The derivation done here yields the best risk estimate, in the MMSE sense, among all linear estimators, and it can be shown that if $\mathbf{x_t}$ is Gaussian distributed, this estimator is the best among all estimators, linear and non-linear included \cite{kalman_optimal}.

\bibliographystyle{ACM-Reference-Format}
\bibliography{ref} 


\begin{thebibliography}{35}

%%% the Web address.

\ifx \showCODEN    \undefined \def \showCODEN     #1{\unskip}     \fi
\ifx \showDOI      \undefined \def \showDOI       #1{#1}\fi
\ifx \showISBNx    \undefined \def \showISBNx     #1{\unskip}     \fi
\ifx \showISBNxiii \undefined \def \showISBNxiii  #1{\unskip}     \fi
\ifx \showISSN     \undefined \def \showISSN      #1{\unskip}     \fi
\ifx \showLCCN     \undefined \def \showLCCN      #1{\unskip}     \fi
\ifx \shownote     \undefined \def \shownote      #1{#1}          \fi
\ifx \showarticletitle \undefined \def \showarticletitle #1{#1}   \fi
\ifx \showURL      \undefined \def \showURL       {\relax}        \fi
% invisible to TeX
\providecommand\bibfield[2]{#2}
\providecommand\bibinfo[2]{#2}
\providecommand\natexlab[1]{#1}
\providecommand\showeprint[2][]{arXiv:#2}

\bibitem[\protect\citeauthoryear{Akinrolabu, Nurse, Martin, and New}{Akinrolabu
  et~al\mbox{.}}{2019}]%
        {cloud_risk_assess}
\bibfield{author}{\bibinfo{person}{Olusola Akinrolabu},
  \bibinfo{person}{Jason~RC Nurse}, \bibinfo{person}{Andrew Martin}, {and}
  \bibinfo{person}{Steve New}.} \bibinfo{year}{2019}\natexlab{}.
\newblock \showarticletitle{Cyber risk assessment in cloud provider
  environments: Current models and future needs}.
\newblock \bibinfo{journal}{\emph{Computers \& Security}}  \bibinfo{volume}{87}
  (\bibinfo{year}{2019}), \bibinfo{pages}{101600}.
\newblock
\urldef\tempurl%
\url{https://doi.org/10.1016/j.cose.2019.101600}
\showDOI{\tempurl}


\bibitem[\protect\citeauthoryear{Albakri, Shanmugam, Samy, Idris, and
  Ahmed}{Albakri et~al\mbox{.}}{2014}]%
        {albakri2014security}
\bibfield{author}{\bibinfo{person}{Sameer~Hasan Albakri},
  \bibinfo{person}{Bharanidharan Shanmugam}, \bibinfo{person}{Ganthan~Narayana
  Samy}, \bibinfo{person}{Norbik~Bashah Idris}, {and} \bibinfo{person}{Azuan
  Ahmed}.} \bibinfo{year}{2014}\natexlab{}.
\newblock \showarticletitle{Security risk assessment framework for cloud
  computing environments}.
\newblock \bibinfo{journal}{\emph{Security and Communication Networks}}
  \bibinfo{volume}{7}, \bibinfo{number}{11} (\bibinfo{year}{2014}),
  \bibinfo{pages}{2114--2124}.
\newblock


\bibitem[\protect\citeauthoryear{Aven}{Aven}{2016}]%
        {risk_analysis}
\bibfield{author}{\bibinfo{person}{Terje Aven}.}
  \bibinfo{year}{2016}\natexlab{}.
\newblock \showarticletitle{Risk assessment and risk management: Review of
  recent advances on their foundation}.
\newblock \bibinfo{journal}{\emph{European Journal of Operational Research}}
  \bibinfo{volume}{253}, \bibinfo{number}{1} (\bibinfo{year}{2016}),
  \bibinfo{pages}{1--13}.
\newblock


\bibitem[\protect\citeauthoryear{Bisong}{Bisong}{2019}]%
        {bisong2019google}
\bibfield{author}{\bibinfo{person}{Ekaba Bisong}.}
  \bibinfo{year}{2019}\natexlab{}.
\newblock \bibinfo{booktitle}{\emph{Building machine learning and deep learning
  models on google cloud platform: A comprehensive guide for beginners}}.
\newblock \bibinfo{publisher}{Apress Berkeley, CA}, \bibinfo{address}{Berkeley,
  CA}. 59--64 pages.
\newblock
\urldef\tempurl%
\url{https://doi.org/10.1007/978-1-4842-4470-8}
\showDOI{\tempurl}


\bibitem[\protect\citeauthoryear{Brown and Hwang}{Brown and Hwang}{1992}]%
        {kalman_optimal}
\bibfield{author}{\bibinfo{person}{Robert~G. Brown} {and}
  \bibinfo{person}{Patrickyc Hwang}.} \bibinfo{year}{1992}\natexlab{}.
\newblock \bibinfo{booktitle}{\emph{Introduction to random signals and applied
  Kalman filtering}}.
\newblock \bibinfo{publisher}{New York, John Wiley \& Sons, Inc., 1992. 512},
  \bibinfo{address}{New York, NY}.
\newblock


\bibitem[\protect\citeauthoryear{Chakravarthia and Ramanathanb}{Chakravarthia
  and Ramanathanb}{2016}]%
        {chakravarthiamonitoring}
\bibfield{author}{\bibinfo{person}{S~Sreenivasa Chakravarthia} {and}
  \bibinfo{person}{Sakkaravarthi Ramanathanb}.}
  \bibinfo{year}{2016}\natexlab{}.
\newblock \showarticletitle{Monitoring Cloud Services Exploitation using
  Heterogeneous Network Flowdata Analysis}.
\newblock \bibinfo{journal}{\emph{International Journal of Control Theory and
  Applications}} \bibinfo{volume}{9}, \bibinfo{number}{51}
  (\bibinfo{year}{2016}), \bibinfo{pages}{299--309}.
\newblock


\bibitem[\protect\citeauthoryear{Chang and Branco}{Chang and Branco}{2021}]%
        {chang2021graph}
\bibfield{author}{\bibinfo{person}{Liyan Chang} {and} \bibinfo{person}{Paula
  Branco}.} \bibinfo{year}{2021}\natexlab{}.
\newblock \bibinfo{title}{Graph-based Solutions with Residuals for Intrusion
  Detection: the Modified E-GraphSAGE and E-ResGAT Algorithms}.
\newblock
\newblock
\showeprint[arxiv]{2111.13597}~[cs.CR]
\urldef\tempurl%
\url{https://arxiv.org/abs/2111.13597}
\showURL{%
\tempurl}


\bibitem[\protect\citeauthoryear{Cisco}{Cisco}{2023}]%
        {cisco_2023}
\bibfield{author}{\bibinfo{person}{Cisco}.} \bibinfo{year}{2023}\natexlab{}.
\newblock \bibinfo{title}{Cisco Secure Network Analytics (Stealthwatch) - Cisco
  Security Analytics {[White paper]}}.
\newblock
\newblock
\urldef\tempurl%
\url{https://www.cisco.com/c/en/us/products/collateral/security/stealthwatch/white-paper-c11-740605.html}
\showURL{%
\tempurl}
\newblock
\shownote{\href{https://www.cisco.com/c/en/us/products/collateral/security/stealthwatch/white-paper-c11-740605.html}{https://www.cisco.com/c/en/us/products/collateral/security/stealthwatch/white-paper-c11-740605.html}.}


\bibitem[\protect\citeauthoryear{Dias and Correia}{Dias and Correia}{2020}]%
        {dias2020big}
\bibfield{author}{\bibinfo{person}{Luis~Filipe Dias} {and}
  \bibinfo{person}{Miguel Correia}.} \bibinfo{year}{2020}\natexlab{}.
\newblock \showarticletitle{Big data analytics for intrusion detection: an
  overview}.
\newblock \bibinfo{journal}{\emph{Handbook of Research on Machine and Deep
  Learning Applications for Cyber Security}} (\bibinfo{year}{2020}),
  \bibinfo{pages}{292--316}.
\newblock


\bibitem[\protect\citeauthoryear{Ergu, Kou, Shi, and Shi}{Ergu
  et~al\mbox{.}}{2014}]%
        {risk_assess1}
\bibfield{author}{\bibinfo{person}{Daji Ergu}, \bibinfo{person}{Gang Kou},
  \bibinfo{person}{Yong Shi}, {and} \bibinfo{person}{Yu Shi}.}
  \bibinfo{year}{2014}\natexlab{}.
\newblock \showarticletitle{Analytic network process in risk assessment and
  decision analysis}.
\newblock \bibinfo{journal}{\emph{Computers \& Operations Research}}
  \bibinfo{volume}{42} (\bibinfo{year}{2014}), \bibinfo{pages}{58--74}.
\newblock


\bibitem[\protect\citeauthoryear{Glymour, Zhang, and Spirtes}{Glymour
  et~al\mbox{.}}{2019}]%
        {glymour_causal}
\bibfield{author}{\bibinfo{person}{Clark Glymour}, \bibinfo{person}{Kun Zhang},
  {and} \bibinfo{person}{Peter Spirtes}.} \bibinfo{year}{2019}\natexlab{}.
\newblock \showarticletitle{Review of causal discovery methods based on
  graphical models}.
\newblock \bibinfo{journal}{\emph{Frontiers in genetics}}  \bibinfo{volume}{10}
  (\bibinfo{year}{2019}), \bibinfo{pages}{524}.
\newblock


\bibitem[\protect\citeauthoryear{Hagen and Kahng}{Hagen and Kahng}{1992}]%
        {spectral}
\bibfield{author}{\bibinfo{person}{Lars Hagen} {and} \bibinfo{person}{Andrew~B
  Kahng}.} \bibinfo{year}{1992}\natexlab{}.
\newblock \showarticletitle{New spectral methods for ratio cut partitioning and
  clustering}.
\newblock \bibinfo{journal}{\emph{IEEE transactions on computer-aided design of
  integrated circuits and systems}} \bibinfo{volume}{11}, \bibinfo{number}{9}
  (\bibinfo{year}{1992}), \bibinfo{pages}{1074--1085}.
\newblock


\bibitem[\protect\citeauthoryear{Hofstede, {\v{C}}eleda, Trammell, Drago,
  Sadre, Sperotto, and Pras}{Hofstede et~al\mbox{.}}{2014}]%
        {hofstede2014flow}
\bibfield{author}{\bibinfo{person}{Rick Hofstede}, \bibinfo{person}{Pavel
  {\v{C}}eleda}, \bibinfo{person}{Brian Trammell}, \bibinfo{person}{Idilio
  Drago}, \bibinfo{person}{Ramin Sadre}, \bibinfo{person}{Anna Sperotto}, {and}
  \bibinfo{person}{Aiko Pras}.} \bibinfo{year}{2014}\natexlab{}.
\newblock \showarticletitle{Flow monitoring explained: From packet capture to
  data analysis with netflow and ipfix}.
\newblock \bibinfo{journal}{\emph{IEEE Communications Surveys \& Tutorials}}
  \bibinfo{volume}{16}, \bibinfo{number}{4} (\bibinfo{year}{2014}),
  \bibinfo{pages}{2037--2064}.
\newblock


\bibitem[\protect\citeauthoryear{Horvath}{Horvath}{2011}]%
        {horvath2011weighted}
\bibfield{author}{\bibinfo{person}{Steve Horvath}.}
  \bibinfo{year}{2011}\natexlab{}.
\newblock \bibinfo{booktitle}{\emph{Weighted network analysis: applications in
  genomics and systems biology}}.
\newblock \bibinfo{publisher}{Springer Science \& Business Media},
  \bibinfo{address}{Los Angeles, CA}.
\newblock


\bibitem[\protect\citeauthoryear{{International Organization for
  Standardization}}{{International Organization for Standardization}}{2018}]%
        {iso31000:2018}
\bibfield{author}{\bibinfo{person}{{International Organization for
  Standardization}}.} \bibinfo{year}{2018}\natexlab{}.
\newblock \bibinfo{title}{{ISO 31000: Risk Management - Guidelines}}.
\newblock
\newblock
\newblock
\shownote{\href{https://www.iso.org/obp/ui\#iso:std:iso:31000:ed-2:v1:en}{https://www.iso.org/obp/ui\#iso:std:iso:31000:ed-2:v1:en}.}


\bibitem[\protect\citeauthoryear{Lallie, Debattista, and Bal}{Lallie
  et~al\mbox{.}}{2020}]%
        {attackgraphreview}
\bibfield{author}{\bibinfo{person}{Harjinder~Singh Lallie},
  \bibinfo{person}{Kurt Debattista}, {and} \bibinfo{person}{Jay Bal}.}
  \bibinfo{year}{2020}\natexlab{}.
\newblock \showarticletitle{A review of attack graph and attack tree visual
  syntax in cyber security}.
\newblock \bibinfo{journal}{\emph{Computer Science Review}}
  \bibinfo{volume}{35} (\bibinfo{year}{2020}), \bibinfo{pages}{100219}.
\newblock


\bibitem[\protect\citeauthoryear{Lashkari, Gil, Mamun, and Ghorbani}{Lashkari
  et~al\mbox{.}}{2017}]%
        {cicflowmeter}
\bibfield{author}{\bibinfo{person}{Arash~Habibi Lashkari},
  \bibinfo{person}{Gerard~Draper Gil}, \bibinfo{person}{Mohammad Saiful~Islam
  Mamun}, {and} \bibinfo{person}{Ali~A. Ghorbani}.}
  \bibinfo{year}{2017}\natexlab{}.
\newblock \showarticletitle{Characterization of Tor Traffic using Time based
  Features}. In \bibinfo{booktitle}{\emph{Proceedings of the 3rd International
  Conference on Information Systems Security and Privacy - Volume 1: ICISSP,}}.
  INSTICC, \bibinfo{publisher}{SciTePress}, \bibinfo{address}{Porto, Portugal},
  \bibinfo{pages}{253--262}.
\newblock
\showISBNx{978-989-758-209-7}
\urldef\tempurl%
\url{https://doi.org/10.5220/0006105602530262}
\showDOI{\tempurl}


\bibitem[\protect\citeauthoryear{Li, Tian, Wu, Cao, Shen, and Long}{Li
  et~al\mbox{.}}{2020}]%
        {li2020cloud}
\bibfield{author}{\bibinfo{person}{Qianmu Li}, \bibinfo{person}{Youhui Tian},
  \bibinfo{person}{Qiang Wu}, \bibinfo{person}{Qi Cao},
  \bibinfo{person}{Haiyuan Shen}, {and} \bibinfo{person}{Huaqiu Long}.}
  \bibinfo{year}{2020}\natexlab{}.
\newblock \showarticletitle{A cloud-fog-edge closed-loop feedback security risk
  prediction method}.
\newblock \bibinfo{journal}{\emph{IEEE Access}}  \bibinfo{volume}{8}
  (\bibinfo{year}{2020}), \bibinfo{pages}{29004--29020}.
\newblock


\bibitem[\protect\citeauthoryear{Maluf, Sudhaakar, and Choo}{Maluf
  et~al\mbox{.}}{2018}]%
        {maluf2018trust}
\bibfield{author}{\bibinfo{person}{David~A Maluf}, \bibinfo{person}{Raghuram~S
  Sudhaakar}, {and} \bibinfo{person}{Kim-Kwang~Raymond Choo}.}
  \bibinfo{year}{2018}\natexlab{}.
\newblock \showarticletitle{Trust Erosion: Dealing with Unknown-Unknowns in
  Cloud Security}.
\newblock \bibinfo{journal}{\emph{IEEE Cloud Computing}} \bibinfo{volume}{5},
  \bibinfo{number}{4} (\bibinfo{year}{2018}), \bibinfo{pages}{24--32}.
\newblock


\bibitem[\protect\citeauthoryear{Nagaraja}{Nagaraja}{2014}]%
        {nagaraja2014botyacc}
\bibfield{author}{\bibinfo{person}{Shishir Nagaraja}.}
  \bibinfo{year}{2014}\natexlab{}.
\newblock \showarticletitle{Botyacc: Unified p2p botnet detection using
  behavioural analysis and graph analysis}. In
  \bibinfo{booktitle}{\emph{European Symposium on Research in Computer
  Security}}. \bibinfo{publisher}{Springer}, \bibinfo{address}{Cham},
  \bibinfo{pages}{439--456}.
\newblock


\bibitem[\protect\citeauthoryear{Ou and Singhal}{Ou and Singhal}{2011}]%
        {ou2011quantitative}
\bibfield{author}{\bibinfo{person}{Xinming Ou} {and} \bibinfo{person}{Anoop
  Singhal}.} \bibinfo{year}{2011}\natexlab{}.
\newblock \bibinfo{booktitle}{\emph{Quantitative security risk assessment of
  enterprise networks}}.
\newblock \bibinfo{publisher}{Springer}, \bibinfo{address}{New York, NY}.
\newblock


\bibitem[\protect\citeauthoryear{Pedregosa, Varoquaux, Gramfort, Michel,
  Thirion, Grisel, Blondel, Prettenhofer, Weiss, Dubourg, Vanderplas, Passos,
  Cournapeau, Brucher, Perrot, and Duchesnay}{Pedregosa et~al\mbox{.}}{2011}]%
        {scikit-learn}
\bibfield{author}{\bibinfo{person}{F. Pedregosa}, \bibinfo{person}{G.
  Varoquaux}, \bibinfo{person}{A. Gramfort}, \bibinfo{person}{V. Michel},
  \bibinfo{person}{B. Thirion}, \bibinfo{person}{O. Grisel},
  \bibinfo{person}{M. Blondel}, \bibinfo{person}{P. Prettenhofer},
  \bibinfo{person}{R. Weiss}, \bibinfo{person}{V. Dubourg}, \bibinfo{person}{J.
  Vanderplas}, \bibinfo{person}{A. Passos}, \bibinfo{person}{D. Cournapeau},
  \bibinfo{person}{M. Brucher}, \bibinfo{person}{M. Perrot}, {and}
  \bibinfo{person}{E. Duchesnay}.} \bibinfo{year}{2011}\natexlab{}.
\newblock \showarticletitle{Scikit-learn: Machine Learning in {P}ython}.
\newblock \bibinfo{journal}{\emph{Journal of Machine Learning Research}}
  \bibinfo{volume}{12} (\bibinfo{year}{2011}), \bibinfo{pages}{2825--2830}.
\newblock


\bibitem[\protect\citeauthoryear{Quinn, Coleman, Kiyavash, and
  Hatsopoulos}{Quinn et~al\mbox{.}}{2011}]%
        {quinn2011estimating}
\bibfield{author}{\bibinfo{person}{Christopher~J Quinn},
  \bibinfo{person}{Todd~P Coleman}, \bibinfo{person}{Negar Kiyavash}, {and}
  \bibinfo{person}{Nicholas~G Hatsopoulos}.} \bibinfo{year}{2011}\natexlab{}.
\newblock \showarticletitle{Estimating the directed information to infer causal
  relationships in ensemble neural spike train recordings}.
\newblock \bibinfo{journal}{\emph{Journal of computational neuroscience}}
  \bibinfo{volume}{30}, \bibinfo{number}{1} (\bibinfo{year}{2011}),
  \bibinfo{pages}{17--44}.
\newblock


\bibitem[\protect\citeauthoryear{Ralston, Graham, and Hieb}{Ralston
  et~al\mbox{.}}{2007}]%
        {ralston2007cyber}
\bibfield{author}{\bibinfo{person}{Patricia~AS Ralston},
  \bibinfo{person}{James~H Graham}, {and} \bibinfo{person}{Jefferey~L Hieb}.}
  \bibinfo{year}{2007}\natexlab{}.
\newblock \showarticletitle{Cyber security risk assessment for SCADA and DCS
  networks}.
\newblock \bibinfo{journal}{\emph{ISA transactions}} \bibinfo{volume}{46},
  \bibinfo{number}{4} (\bibinfo{year}{2007}), \bibinfo{pages}{583--594}.
\newblock


\bibitem[\protect\citeauthoryear{Ramos, Lazar, Holanda~Filho, and
  Rodrigues}{Ramos et~al\mbox{.}}{2017}]%
        {ramos2017model}
\bibfield{author}{\bibinfo{person}{Alex Ramos}, \bibinfo{person}{Marcella
  Lazar}, \bibinfo{person}{Raimir Holanda~Filho}, {and}
  \bibinfo{person}{Joel~JPC Rodrigues}.} \bibinfo{year}{2017}\natexlab{}.
\newblock \showarticletitle{Model-based quantitative network security metrics:
  A survey}.
\newblock \bibinfo{journal}{\emph{IEEE Communications Surveys \& Tutorials}}
  \bibinfo{volume}{19}, \bibinfo{number}{4} (\bibinfo{year}{2017}),
  \bibinfo{pages}{2704--2734}.
\newblock


\bibitem[\protect\citeauthoryear{Saripalli and Walters}{Saripalli and
  Walters}{2010}]%
        {quirc}
\bibfield{author}{\bibinfo{person}{Prasad Saripalli} {and} \bibinfo{person}{Ben
  Walters}.} \bibinfo{year}{2010}\natexlab{}.
\newblock \showarticletitle{Quirc: A quantitative impact and risk assessment
  framework for cloud security}. In \bibinfo{booktitle}{\emph{2010 IEEE 3rd
  international conference on cloud computing}}. \bibinfo{publisher}{IEEE},
  \bibinfo{address}{Miami, FL}, \bibinfo{pages}{280--288}.
\newblock


\bibitem[\protect\citeauthoryear{Sendi and Cheriet}{Sendi and Cheriet}{2014}]%
        {sendi_cloud}
\bibfield{author}{\bibinfo{person}{Alireza~Shameli Sendi} {and}
  \bibinfo{person}{Mohamed Cheriet}.} \bibinfo{year}{2014}\natexlab{}.
\newblock \showarticletitle{Cloud Computing: A Risk Assessment Model}. In
  \bibinfo{booktitle}{\emph{2014 IEEE International Conference on Cloud
  Engineering}}. \bibinfo{publisher}{IEEE}, \bibinfo{address}{Boston, MA},
  \bibinfo{pages}{147--152}.
\newblock
\urldef\tempurl%
\url{https://doi.org/10.1109/IC2E.2014.17}
\showDOI{\tempurl}


\bibitem[\protect\citeauthoryear{Sharafaldin, Habibi~Lashkari, and
  Ghorbani}{Sharafaldin et~al\mbox{.}}{2018a}]%
        {cicids2017_dataset}
\bibfield{author}{\bibinfo{person}{Iman Sharafaldin}, \bibinfo{person}{Arash
  Habibi~Lashkari}, {and} \bibinfo{person}{Ali Ghorbani}.}
  \bibinfo{year}{2018}\natexlab{a}.
\newblock \bibinfo{title}{Intrusion detection evaluation dataset {(CIC-IDS2017)
  [Dataset]}}.
\newblock
\newblock
\newblock
\shownote{\href{http://www.unb.ca/cic/datasets/IDS2017.html}{http://www.unb.ca/cic/datasets/IDS2017.html}.}


\bibitem[\protect\citeauthoryear{Sharafaldin, Habibi~Lashkari, and
  Ghorbani}{Sharafaldin et~al\mbox{.}}{2018b}]%
        {cicids2017}
\bibfield{author}{\bibinfo{person}{Iman Sharafaldin}, \bibinfo{person}{Arash
  Habibi~Lashkari}, {and} \bibinfo{person}{Ali Ghorbani}.}
  \bibinfo{year}{2018}\natexlab{b}.
\newblock \showarticletitle{Toward Generating a New Intrusion Detection Dataset
  and Intrusion Traffic Characterization}. In \bibinfo{booktitle}{\emph{4th
  International Conference on Information Systems Security and Privacy (ICISSP
  2018)}}. \bibinfo{publisher}{ICISSP}, \bibinfo{address}{Madeira, Portugal},
  \bibinfo{pages}{108--116}.
\newblock
\urldef\tempurl%
\url{https://doi.org/10.5220/0006639801080116}
\showDOI{\tempurl}


\bibitem[\protect\citeauthoryear{Simon}{Simon}{2006}]%
        {kalman_book}
\bibfield{author}{\bibinfo{person}{Dan Simon}.}
  \bibinfo{year}{2006}\natexlab{}.
\newblock \bibinfo{booktitle}{\emph{Optimal state estimation: Kalman, H
  infinity, and nonlinear approaches}}.
\newblock \bibinfo{publisher}{John Wiley \& Sons}, \bibinfo{address}{Hoboken,
  NJ}.
\newblock


\bibitem[\protect\citeauthoryear{Sperotto, Schaffrath, Sadre, Morariu, Pras,
  and Stiller}{Sperotto et~al\mbox{.}}{2010}]%
        {sperotto2010overview}
\bibfield{author}{\bibinfo{person}{Anna Sperotto}, \bibinfo{person}{Gregor
  Schaffrath}, \bibinfo{person}{Ramin Sadre}, \bibinfo{person}{Cristian
  Morariu}, \bibinfo{person}{Aiko Pras}, {and} \bibinfo{person}{Burkhard
  Stiller}.} \bibinfo{year}{2010}\natexlab{}.
\newblock \showarticletitle{An overview of IP flow-based intrusion detection}.
\newblock \bibinfo{journal}{\emph{IEEE communications surveys \& tutorials}}
  \bibinfo{volume}{12}, \bibinfo{number}{3} (\bibinfo{year}{2010}),
  \bibinfo{pages}{343--356}.
\newblock


\bibitem[\protect\citeauthoryear{Wang, Alahmadi, Zhu, and Li}{Wang
  et~al\mbox{.}}{2015}]%
        {wang2015brain}
\bibfield{author}{\bibinfo{person}{Zhe Wang}, \bibinfo{person}{Ahmed Alahmadi},
  \bibinfo{person}{David Zhu}, {and} \bibinfo{person}{Tongtong Li}.}
  \bibinfo{year}{2015}\natexlab{}.
\newblock \showarticletitle{Brain functional connectivity analysis using mutual
  information}. In \bibinfo{booktitle}{\emph{2015 IEEE Global Conference on
  Signal and Information Processing (GlobalSIP)}}. \bibinfo{publisher}{IEEE},
  \bibinfo{address}{Orlando, FL}, \bibinfo{pages}{542--546}.
\newblock


\bibitem[\protect\citeauthoryear{Wu, Wang, Huang, and Zhang}{Wu
  et~al\mbox{.}}{2022}]%
        {wu2022intrusion}
\bibfield{author}{\bibinfo{person}{Junrui Wu}, \bibinfo{person}{Wenyong Wang},
  \bibinfo{person}{Lisheng Huang}, {and} \bibinfo{person}{Fengjun Zhang}.}
  \bibinfo{year}{2022}\natexlab{}.
\newblock \showarticletitle{Intrusion detection technique based on flow
  aggregation and latent semantic analysis}.
\newblock \bibinfo{journal}{\emph{Applied Soft Computing}}
  \bibinfo{volume}{127} (\bibinfo{year}{2022}), \bibinfo{pages}{109375}.
\newblock


\bibitem[\protect\citeauthoryear{Young, Neveu, Byrne, and Aazhang}{Young
  et~al\mbox{.}}{2021}]%
        {young2021inferring}
\bibfield{author}{\bibinfo{person}{Joseph Young}, \bibinfo{person}{Curtis~L
  Neveu}, \bibinfo{person}{John~H Byrne}, {and} \bibinfo{person}{Behnaam
  Aazhang}.} \bibinfo{year}{2021}\natexlab{}.
\newblock \showarticletitle{Inferring functional connectivity through graphical
  directed information}.
\newblock \bibinfo{journal}{\emph{Journal of Neural Engineering}}
  \bibinfo{volume}{18}, \bibinfo{number}{4} (\bibinfo{year}{2021}),
  \bibinfo{pages}{046019}.
\newblock


\bibitem[\protect\citeauthoryear{Zhou, Guo, Huang, Guo, and Zhu}{Zhou
  et~al\mbox{.}}{2016}]%
        {tvmm}
\bibfield{author}{\bibinfo{person}{Chao Zhou}, \bibinfo{person}{Yajuan Guo},
  \bibinfo{person}{Wei Huang}, \bibinfo{person}{Jing Guo}, {and}
  \bibinfo{person}{Daohua Zhu}.} \bibinfo{year}{2016}\natexlab{}.
\newblock \showarticletitle{Network Security Risk Prediction Based on
  Time-Varying Markov Model}. In \bibinfo{booktitle}{\emph{Proceedings of the
  2016 4th International Conference on Mechanical Materials and Manufacturing
  Engineering}}. \bibinfo{publisher}{Atlantis Press}, \bibinfo{address}{Wuhan,
  China}, \bibinfo{pages}{212--215}.
\newblock
\showISBNx{978-94-6252-221-3}
\showISSN{2352-5401}
\urldef\tempurl%
\url{https://doi.org/10.2991/mmme-16.2016.49}
\showDOI{\tempurl}


\end{thebibliography}

\end{document}